\documentclass{aa}  
\usepackage{graphicx}
\usepackage{caption}
\usepackage{subcaption}
\usepackage[varg]{txfonts}
\usepackage{setspace}
\usepackage{booktabs}
\usepackage{multirow, makecell}
\usepackage{float}
\usepackage{mydblfloatfix} 

\usepackage{float}
\usepackage{tikz}
\usepackage{color}
\graphicspath{{./}{./figure/}}

\usepackage{xcolor}

\titlerunning{Helicity and free energy as tools to probe eruptions in
two differently evolving active regions}

\usetikzlibrary{arrows,positioning} 
\tikzset{
    >=stealth',
    punkt/.style={
           rectangle,
           rounded corners,
           draw=black, very thick,
           text width=6.5em,
           minimum height=2em,
           text centered},
    pil/.style={
           ->,
           thick,
           shorten <=2pt,
           shorten >=2pt,}
}

\newcommand\rectagular[1][red]{\begin{tikzpicture}
\draw [fill=red,red] (0.2,0.2) rectangle (0.3,0.3); 
\end{tikzpicture}
}\usepackage{lipsum}

\begin{document}

    \title{Magnetic Helicity and Free Magnetic Energy as Tools to Probe Eruptions in two Differently Evolving Solar Active Regions}
    
    \author{E. Liokati\inst{\ref{inst1}} \and A. Nindos\inst{\ref{inst1}} \and M. K. Georgoulis\inst{\ref{inst2}} }
    
  \institute{Section of Astrogeophysics, Department of Physics, University of Ioannina, 45110, Greece.\label{inst1} \and Research Center for Astronomy and Applied Mathematics, Academy of Athens, Athens, 11527, Greece\label{inst2} }
    
\date{Received date /
Accepted date }

\abstract {} 
{We study the role of magnetic helicity and free magnetic energy in 
the initiation of eruptions in two differently evolving solar active 
regions (ARs).} 
{Using vector magnetograms from the Helioseismic and Magnetic Imager on
board the Solar Dynamics Observatory and a magnetic connectivity-based 
method, we 
calculate the instantaneous relative magnetic helicity and free magnetic
energy budgets for several days in two ARs, AR11890 and AR11618, both 
with complex photospheric magnetic field configurations.} 
{The ARs produced several major eruptive flares while their photospheric
magnetic field exhibited different evolutionary patterns; primarily flux 
decay in AR11890 and primarily flux emergence in AR11618.
Throughout much of their evolution both ARs featured substantial budgets of 
free magnetic energy and of both positive (right-handed) and negative 
(left-handed) helicity. In fact, the 
imbalance between the signed components of their helicity was as low as in the
quiet Sun and their net helicity eventually changed sign 14-19 hours after their
last major flare. Despite such incoherence, the eruptions occurred at times
of net helicity peaks that were co-temporal with peaks in the free 
magnetic energy. 
The percentage losses, associated with the eruptive flares, in the
normalized free magnetic energy were significant, in the range $\sim$10-60\%.
For the magnetic helicity, changes ranged from $\sim$25\% to the removal of the
entire excess helicity of the prevailing sign, leading a roughly zero net
helicity, but with significant equal and opposite budgets of both helicity
senses. Respective values ranged from $(0.3-2) \times 10^{32}$ erg and $(1.3-20)
\times 10^{42}$ Mx$^2$ for energy and helicity losses.
The removal of the slowly varying background component of the free 
energy and helicity (either the net helicity or the prevailing signed component of helicity) timeseries revealed that all eruption-related peaks of 
both quantities exceeded the 2$\sigma$ levels of their detrended timeseries 
above the removed background. There was no eruption when only one or 
none of these quantities exceeded its 2$\sigma$ level.}
{Our results indicate that differently evolving ARs may produce major eruptive flares even when, in addition to the accumulation of significant free
magnetic energy budgets, they accumulate large amounts of both left- and 
right-handed helicity without a strong dominance of one handedness over 
the other. In most cases, these excess budgets appear as localized peaks,
co-temporal with the flare peaks,
in the timeseries of free magnetic energy and helicity (and normalized values
thereof). The corresponding normalized free magnetic energy and helicity 
losses can be very significant at times.} 
\maketitle

\section{Introduction} \label{sec: introduction}

Coronal mass ejections (CMEs) are large-scale expulsions of magnetized
plasma from the solar corona into the interplanetary medium, observed with white-light coronagraphs. Flares are sudden flashes of radiation across virtually the entire
electromagnetic spectrum.
In contrast to flares that usually occur in active regions (ARs), CMEs can occur both in ARs and away of ARs. 
Not all flares are accompanied by CMEs but all active-region CMEs are 
accompanied by flares. Due to the lack of sufficient magnetic energy, quiet-Sun
CMEs are statistically slower and are not accompanied by major flares
\citep[e.g.][]{Webb_1987, Sheeley_1983, St.Cyr_1991, Harrison_1995, Andrews_2003}.
When they do, the flares are called eruptive, otherwise
they are called confined. A close temporal correlation and synchronization
has been reported in several cases of paired flare-CME events
\citep[e.g.][]{Zhang_2001, Zhang_2004, Gallagher_2003, Vrsnak_2004, Yashiro_2006,Maricic_2007, Vrsnak_2008, Temmer_2008, Temmer_2010, Schmieder_2015,Gou_2020}. In the strongest events, practically always flares and CMEs occur together \citep[e.g.][]{Yashiro_2006}. 

Flares and CMEs occur in regions where a significant build-up of electric currents has stressed the magnetic field which, as a result, deviates from the potential state \citep[e.g. see the reviews by][and references therein]{Forbes_2000, Klimchuk_2001, Aulanier_2014, Schmieder_2015, Cheng_2017, Green_2018, Georgoulis_2019}. 
CMEs may result from the catastrophic loss of equilibrium between the magnetic pressure and tension acting on such regions. 
Magnetic pressure is favored in structures of strong magnetic field that tend to expand into areas of weak magnetic field, whereas magnetic tension acts as a restraining agent keeping the stressed magnetic structure contained or strapped by the overlying coronal magnetic field. Magnetic confinement fails, and thus a CME is generated, either due to magnetic reconnection or due to the prevelance of some ideal instability that develops when a previously confined magnetic 
structure is enabled (by means of magnetic helicity and/or energy) to 
initiate 
an outward expansion against the overlying background magnetic field. 
In the former case, the pre-eruptive configuration is likely a sheared magnetic arcade, that is, sets of loops whose planes deviate significantly from the local normal to the magnetic polarity inversion line (PIL). 
Examples include the models by e.g. \citet{Sturrock_1966, Antioxos_Devore_Klimchuk_1999, Fan_2001, Manchester_2003, MacNeice_2004, Lynch_2008,van_der_Holst_2009, Fang_2010, Fang_2012} which have occssionally derived support from observations \citep[e.g.][]{Aulanier_2000, Ugarte-Urra_2007}. 
In the latter case, the pre-eruptive configuration is likely a magnetic flux rope, that is a set of magnetic field lines winding about an axial field line in an organized manner. Relevant models include those developed by \citet{Amari_2000, Amari_2004, Amari_2005, Torok_kliem_2005, Kleim_torok_2006, Fan_2007, Archontis_Torok_2008, Archontis_Hood_2012}. Several studies report observational support in favor of eruptions that involve preexisting flux ropes \citep[e.g.][]{Green_Kliem_2009, Zhang_cheng_ding_2012, Patsourakos_2013, Cheng_2013, vourlidas2014flux, Nindos_flux_rope_2020}.

There are several patterns of magnetic field evolution that may lead to 
stressed magnetic configurations required for the initiation of CMEs. 
These inlude: (1) magnetic flux emergence, in which vertical motions transfer current-carrying magnetic flux from the interior to the atmosphere of the Sun \citep[e.g.][]{Fan_2009, Archontis_2012, Hood_2012, Toriumi_2014, Archontis_2019}, (2) PIL-aligned shearing motions of the photospheric magnetic field \citep[e.g.][]{Zhang_1995, Demoulin_2002, Nindos_zhang_2002, Georgoulis_Titov_2012, Vemareddy_2017, Vemareddy_2019}, and (3) magnetic flux cancellation in which small-scale opposite magnetic polarities converge, interact via magnetic reconnection, and then subsequently submerge into the solar interior along the PIL \citep{Babcock_1955, Martin_1985, vanballegooijen_1989,Green_2011, yardley_2018}. We note that these processes may appear independently or in tandem in regions that will subsequently erupt.  
\citet{Georgoulis_Titov_2012} \citep[see also][]{Georgoulis_2019} have 
discussed a scenario in which the action of the Lorentz force along strong PILs
that is eventually triggered by the development of intense non-neutralized 
currents could account for the velocity shear as long as flux emergence takes 
place. Furthermore, \citet{Chintzoglou_2019} proposed a mechanism of so-called 
``collisional shearing'' between two emerging flux tubes that could account for
all three mechanisms. 

In stressed magnetic configurations 
the most important term of the magnetic energy is the so-called free 
energy, that is exclusively due to electric currents. It is only this term that
can be extracted (via elimination of currents) and converted to other energy 
forms
\citep[e.g. see][]{Priest_book_2014}. Another quantity that is often used for 
the description of non-potential magnetic fields is magnetic helicity which is
a signed quantity that quantifies the twist, writhe and linkage of a set of 
magnetic flux tubes 
\citep[e.g. see the review by][]{Pevtson_2014}. In ideal plasmas, magnetic
helicity is perfectly conserved 
\citep[e.g. see][]{Sturrock_1994} while in 
magnetic reconnection and other nonideal processes, it is very well conserved 
if the plasma magnetic Reynolds number is high 
\citep[e.g. see][]{berger1984PhDthesis,Berger_1999,pariat2015}. Free 
magnetic energy is 
released in the course of flares, CMEs and smaller-scale dissipative events 
(e.g. subflares, jets) while helicity can either be removed by CMEs or be
transferred during reconnection events to larger scales via existing magnetic 
connections.

The role of free magnetic energy in the initiation of solar eruptions is widely known \citep[e.g.][]{Neukirch_2005, Schrijver2009} but the role of helicity has been debated as some theoretical investigations have demonstrated that helicity is not necessary for CME initiation \citep{MacNeice_2004, Phillips2005, Zuccarello2009}. On the other hand in other theoretical works it is conjectured that
the corona expells excess helicity primarily through CMEs 
\citep[e.g. see][]{Low_1994Magnetohydrodynamic,Low_1996,Zhang_low_2005,Georgoulis_2019}. 
The arguments for such conclusion are as follows. Differential rotation and 
subsurface dynamos constantly generate negative magnetic helicity in the 
northern solar hemisphere and positive magnetic helicity in the southern
hemisphere \citep{Seehafer_1990, Pevtsov_1995}, and this trend does not change
from solar cycle to solar cycle \citep[][]{Pevtsov_2001}. Due to the conserved 
nature of helicity, this process would constantly charge the corona with 
helicity. Furthermore there are no observations showing any significant 
cancellation of helicity across the equator. 
In addition, returning atmospheric helicity back to the solar interior with flux submergence would violate the entropy principle, i.e., result in situations of less entropy than before.
Therefore CMEs appear as the
obvious valves that relieve the Sun from its excess helicity. This conjecture has
been quantified by \citet{Zhang_2006,Zhang_2012} who found that upper limits for
the accumulation of helicity exist which, if crossed, a nonequilibrium state 
develops that may yield a CME. Furthermore, it has been proposed 
\citep[][]{kusano2003,Kusano_2004} that the accumulation of similar budgets of
positive and negative helicity may enable reconnection leading to eruptions.

Observational support for the importance of helicity in the initiation of
solar eruptions include the works by 
\citet{nindos_andrews2004,Labonte_2007,Park_2008,Park_2010,Nindos_2012}. Using 
different methods, \citet{Tziotziou_2012} and \citet{Liokati_2022} have found 
thresholds for both the magnetic helicity ($0.9-2 \times 10^{42}$ Mx$^2$) and 
the free magnetic energy or total magnetic energy ($0.4-2 \times 10^{32}$ erg) 
which, if exceeded, the host AR is likely to erupt. Some authors 
\citep[][]{Pariat_2017,Thalmann_2019,Gupta_2021} advocate that the ratio of
the helicity associated with the current-carrying magnetic field to the total
helicity is a reliable proxy for solar eruptions while both the total helicity
and the magnetic energy are not. \citet{Price_2019} suggest that for
the prediction of eruptive flares the above helicity ratio should be considered
in combination with the free magnetic energy. Interested readers are referred
to the review by \citet{Toriumi_2022} for a comprehensive outlook of our
current understanding of the role of helicity in the occurrence of flares and
CMEs.

Several methods of magnetic helicity estimation 
\citep[for a comparison, see][]{Thalmann_2021} have been developed which include (i) finite-volume methods \citep[see][for a review and comparison of several
implementations of the method]{Valori_2016}, (ii) the connectivity-based method developed by \citet{georgoulis2012magnetic} (see Sect. 3.1 for details), (iii) the helicity-flux integration method \citep[][; see Sect. 3.2 for details]{Chae_2001, Nindos_2003, Pariat_2005, georgoulis_labonte_2007, Liu_2012, dalmasse2014, dalmasse2018}, and (iv) the twist-number method \citep{Guo_2010, Guo_2017}. 
Methods (i), (ii), and (iv) yield the instantaneous helicity 
but in method (iv) only the twist contribution to the helicity is 
calculated. With the flux integration method we obtain only the helicity 
injection rate and thus the helicity change over certain time intervals.

In this paper we study the evolution of helicity and free magnetic energy, as 
quantified by their instantaneous values which are tracked for several days, 
in two eruptive ARs with significantly different magnetic flux evolution. 
Using the connectivity-based
method we show that both the magnetic helicity and the free magnetic 
energy play an important role in the development of eruptions in both ARs. 
In the next section we describe our data base and in Sect. 3 the methods we 
used for the calculation of the magnetic helicity and energy. In Sect. 4
we study the long-term evolution of the free magnetic energy and helicity from
the connectivity-based method. These results are then compared with the 
results
from the flux-integration method. In Sect. 5 we discuss the helicity and free
magnetic energy budgets of the major eruptive flares that occurred in the ARs.
The conclusions and a summary of our work are presented in Sect. 6.

\section{Observations} 

We study two ARs, namely, NOAA AR11890 and 11618. Both showed complex 
photospheric magnetic field configurations that, however,
exhibited different evolution 
patterns. The evolution of the former was dominated primarily by 
magnetic flux 
decay for more than half of the interval that we studied while the 
evolution of the latter was dominated primarily by magnetic flux 
emergence. 
Both ARs produced several major eruptive flares during their passage 
from the earthward solar disk. 

For our study we used vector magnetograms \citep[][]{Hoeksema_hmi_2014} 
from the Helioseismic and Magnetic Imager \citep[HMI;][]{scherrer2012_HMI,Schou_2012} telescope on board the Solar Dynamics Observatory 
\citep[SDO;][]{pesnell2012_SDO}. In particular we employed series of the 
so-called 
HMI.SHARP\_CEA\_720s data products \citep{bobra2014helioseismic} which yield the 
photospheric magnetic field vector in Lambert cylindrical equal-area (CEA) projection.
In these data the vector magnetic field output from the inversion code has been
transformed into spherical heliographic components $B_r$, $B_{\theta}$, and 
$B_{\phi}$ \citep[e.g. see][]{Gary_1990} which are directly related to the
Cartesian heliographic components of the magnetic field via 
[$B_r$, $B_{\theta}$, $B_{\phi}$]
$=$ [$B_z$, $-B_y$, $B_x$] \citep[e.g. see][]{Sun_2013}, where $x$, $y$, and 
$z$ denote solar westward, northward, and vertical directions, respectively. 

The angular resolution of the CEA magnetic field images is 0.03 CEA degrees 
which is equivalent to approximately 360 km per pixel at disk center. The
cadence of our vector field image cubes was 12 min. In Table 1 we show the
start and end times of the observations of the two ARs together with their
corresponding locations on the solar disk. We note that there was a data gap
in HMI observations of AR11618 from 22 November 2012 23:10 UT until 23 November
2012 23:22 UT. 

\begin{table*}
\begin{center}
\caption{HMI observations}
\begin{tabular}{lcccc}
\toprule
NOAA AR & Start  time & Start location  & End time  & End location  \\ 
11890   & 5 November 2013 14:00 & S09E41 & 11 November 2013 15:36 & S12W37   \\ 
11618 & 18 November 2012 20:00  &  N09E39  & 25 November 2012 11:12 & N08W47 \\
\bottomrule
\end{tabular}
\end{center}
\end{table*}

For the recording of flares associated with our ARs we used (1) data from 
NOAA's Geostationary Operation Environmental Satellite (GOES) flare 
catalog\footnote{See \url{https://www.ngdc.noaa.gov/stp/space-weather/solar-data/solar-features/solar-flares/x-rays/goes/xrs/}} 
and (2) images from the Atmospheric Imaging Assembly \citep[AIA;][]{aia_2012_lemen, boerner_2012_aia} telescope onboard SDO at 131~\AA\ and 171~\AA. 
For the detection of the CMEs produced by our ARs we used 
(1) movies from data obtained by the Large Angle and Spectrometric Coronagraph \citep[LASCO;][]{Brueckner_lasco_1955} onboard the Solar and Heliospheric Observatory (SOHO)
that can be found in the Coordinated Data Analysis Workshop (CDAW) SOHO/LASCO CME catalog,\footnote{See \url{https://cdaw.gsfc.nasa.gov/CME_list/}}
\citep[][]{gopalswamy2009_cme_catalog} and 
(2) 211 \AA~AIA/SDO difference images. This particular AIA passband was chosen because it shows better CME-associated dimming regions which were used, together with the presence of ascending loops, as proxies for locating the CME sources.

During the time intervals we studied, several major flares occurred in
both ARs; six in AR11890 (three X-class and three M-class) and four M-class ones
in AR11618. All of these major flares were eruptive.  Furthermore, several
C-class flares occurred during the observations (19 in AR11890 and 7 in AR11618)
which were all confined.

\section{Calculations of magnetic helicity and energy budgets}

\subsection{Connectivity-based method}

For each AR the instantaneous relative magnetic helicity and free magnetic energy 
budgets were computed using the connectivity-based (CB) method developed by 
\citet{georgoulis2012magnetic} who generalized the linear force-free (LFF) 
method of \citet{georgoulis_labonte_2007} into a nonlinear force-free 
(NLFF) one, at the same time incorporating the properties of mutual
helicity as discussed by \citet{demoulin_2006}. This method requires a single 
vector magnetogram whose flux distribution is partitioned. Then a connectivity
matrix containing the magnetic flux associated with connections between 
positive polarity and negative polarity partitions is computed. This computation
is performed with a simulated annealing method which prioritizes connections
between opposite polarity partitions while globally minimizing the connection
lengths. The collection of connections provided by the connectivity matrix is
treated as an ensemble of $N$ slender force-free flux tubes, each with
known footpoint locations, magnetic flux, and force-free parameter. 

The free magnetic energy, $E_f$, and magnetic helicity, $H$, for these flux 
tubes are provided as algebraic sums of a self term ($E_{f,self}$ or $H_{self}$) 
corresponding to the twist and writhe of each flux tube, and a mutual term 
($E_{f,mut}$ or $H_{mut}$) corresponding to interactions between different tubes:

\begin{equation}  \label{equ: energy}
\begin{split}
E_{f} & = E_{f_{self}} + E_{f_{mut}} \\
      & = A d^{2} \sum_{l=1}^{N} \alpha_{l}^{2} \Phi_{l}^{2\lambda} + \frac{1}{8\pi} \sum_{l=1}^{N} \sum_{m=1, l\neq m}^{N} \alpha_{l} \mathcal{L}_{lm}^{arch} \Phi_{l} \Phi_{m} 
\end{split}
\end{equation}

\begin{equation} \label{equ:helicity}
\begin{split}
H & = H_{self} + H_{mut} \\
      & =8 \pi A d^{2}\sum_{l=1}^{N} \alpha_{l} \Phi_{l}^{2 \lambda}+ \sum _{l=1}^{N} \sum_{m=1, l\neq m}^{N} \mathcal{L}_{lm}^{arch} \Phi_{l} \Phi_{m}
\end{split}
\end{equation}
where $\mathit{d}$ is the pixel size, $\mathit{A} $ and $\mathit{\lambda}$ are 
known fitting constants, $l$ and $m$ are different flux tubes with known 
unsigned flux, $ \Phi $, and force-free parameter, $ \alpha $. 
$ \mathcal{L}_{lm}^{arch} $ is a mutual-helicity parameter describing the 
interaction of two arch-like flux tubes that do not wind around each other's
axes \citep[see][]{demoulin_2006}. Therefore the computed free energy and 
helicity can be considered lower limits of their actual value since the 
winding of different 
flux tubes is ignored. We also note that the discrete nature of the CB method
enables the independent computation not only of the self and mutual helicity
and free energy terms, but also of the right-handed (positive) and
left-handed (negative) contributions to the total helicity.

In our computations, care was exercised not to include pixels with negligible  
contributions to the helicity and free energy budgets which could nevertheless 
significantly add to the required computing time. Such pixels may be associated
with quiet-Sun or weak-field regions consisting of numerous
small-scale structures. To this  
end, we used the following thresholds for partitioning the magnetograms of both
ARs: (1)  50 G in $|B_z|$, (2) a minimum magnetic flux of $5 \times 10^{19}$ Mx 
per partition, and (3) a minimum number of 30 pixels per partition. For 
further analysis we only used those partitions that satisfied all of the above 
threshold  criteria. The threshold values we chose satisfied the following 
requirements:  (1) a significant majority of the unsigned magnetic flux should 
be included in  the flux partitioning, (2) the value of a given threshold 
should not change  throughout the evolution of both ARs, and (3) for each AR, 
the thresholds were first tested to images featuring the most dispersed 
magnetic flux distribution and then to  increasingly compact magnetic 
configurations to make sure that the required  computing time is always kept at
a reasonable level. 

\begin{figure*}[]
\centering
\includegraphics[width=15cm]{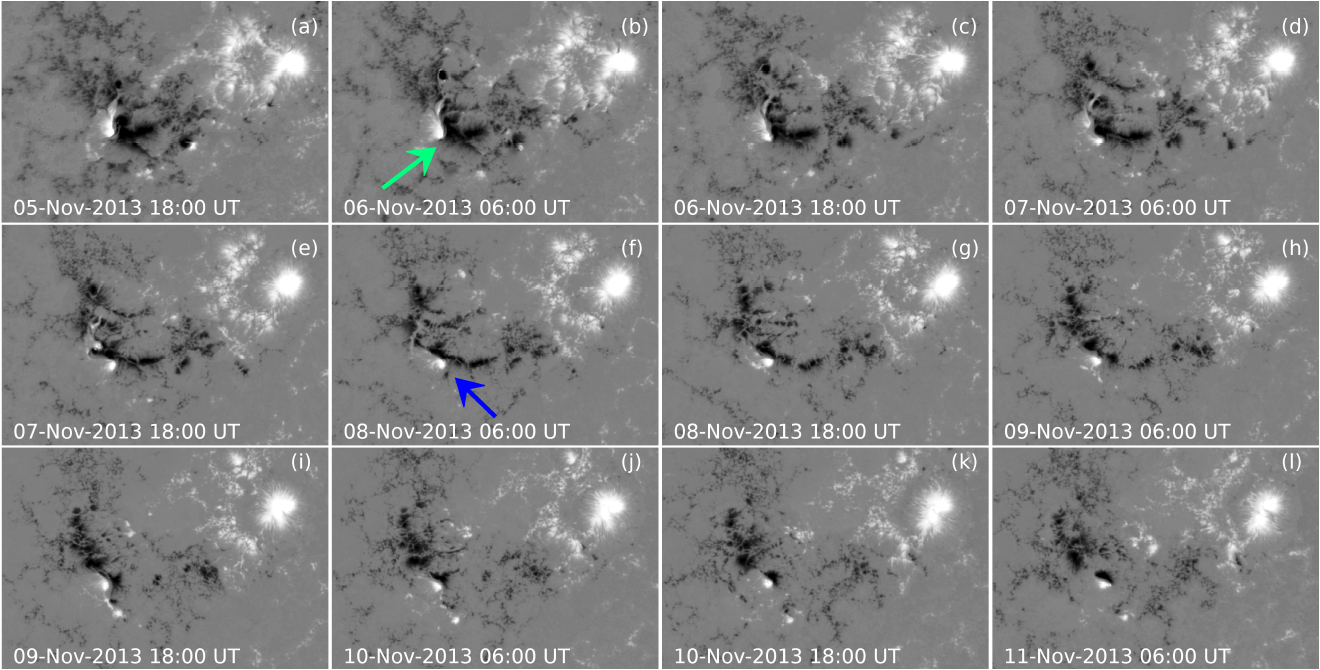}
\caption{Selected images of the normal component of the photospheric magnetic 
field of AR11890 taken by the HMI during the interval studied in this paper. 
The green arrow marks the area where intense magnetic flux decay 
occurred while the blue arrow shows the location of the parasitic
positive polarity at a later stage during the flux decay episode.
The field of view of each panel is $571\arcsec \times 387\arcsec$.}
\label{fig:hmi_11890}
\end{figure*}

\begin{figure*}[]
\centering
\includegraphics[width=15cm]{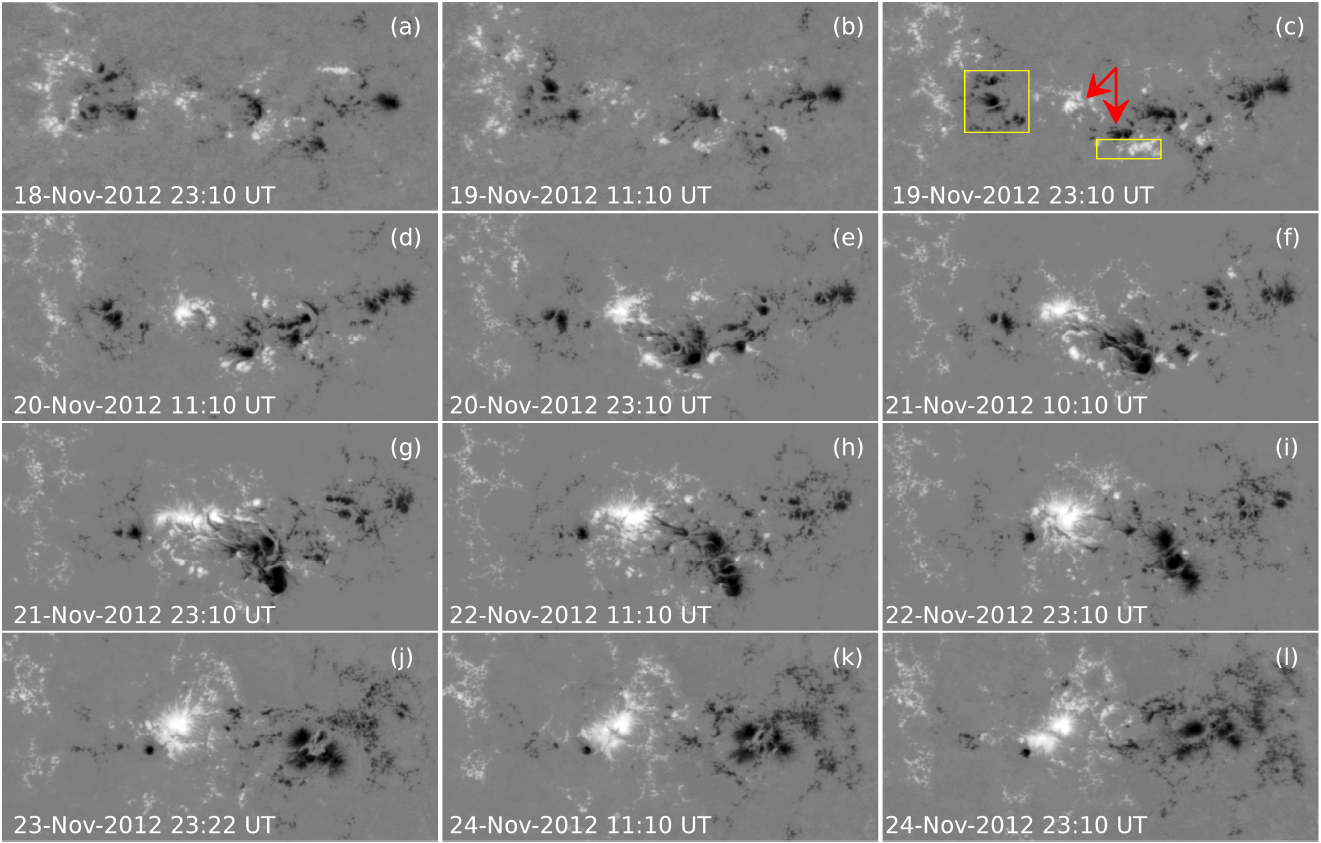}
\caption{Same as Fig. \ref{fig:hmi_11890} but for AR11618. The red 
arrows indicate sites of major flux emergence while the yellow boxes enclose 
areas whose magnetic flux was affected by cancellation. The field of view of
each panel is $611\arcsec \times 291\arcsec$.}
\label{fig:hmi_11618}
\end{figure*}

The uncertainties of the CB-method results have been discussed by 
\citet{georgoulis2012magnetic}. They are usually rather small and 
for this reason we did not use them; instead, we use the standard
deviations of the moving five-point (48-minute) averages of the $H$ and $E_f$
curves, that tend to represent more sizable uncertainties
\citep[see also][]{Moraitis_2021}. The uncertainties of all quantities
that are produced from either $H$ or $E_f$ (see Sect. 4.3, 4.4 and 5) were also
calculated by evaluating the standard deviations of their moving five-point
averages.

\subsection{Helicity and energy flux integration method}

The formulas for the magnetic helicity and magnetic energy fluxes across the
photospheric surface, $S$ are

\begin{equation} \label{equ:helicity flux}
\dfrac{dH}{dt}\Bigg|_{S} = 2 \int_{S} (\mathbf{A}_{P} \cdot \mathbf{B}_{t}) V_{\perp n} dS - 2 \int_{S}(\mathbf{A}_{P} \cdot \mathbf{V}_{\perp t})B_{n}dS
\end{equation}

\begin{equation} \label{equ:energy flux}
\frac{dE}{dt}\Bigg|_{S} = \frac{1}{4\pi} \int_{S} B^{2}_{t} V_{\perp n} dS -\frac{1}{4\pi} \int_{S} (\mathbf{B}_{t} \cdot \mathbf{V}_{\perp t}) B_{n} dS 
\end{equation} 
\citep[][]{berger1984PhDthesis,Berger_1999,Kusano_2002}
where $ \mathbf{A}_{P} $ is the 
vector potential of the potential magnetic field $\mathbf{B}_{P}$, 
$\mathbf{B}_{n}$ and $\mathbf{B}_{t}$ are the normal and tangential components
of the photospheric magnetic field, and $V_{\perp n}$ and $\mathbf{V}_{\perp t}$
are the normal and tangential components of the velocity $V_{\perp}$ which is
perpendicular to the magnetic field lines (cross-field velocity). The 
first terms of Eqs. (3) and (4)
correspond to the contribution from magnetic flux emergence while the second
terms correspond to the contribution from photospheric shuffling. 

The velocity field involved in Eqs. (3) and (4) was computed with the 
Differential Affine Velocity Estimator for Vector Magnetograms 
\citep[DAVE4VM;][]{Schuck_2008} algorithm, applied to sequential pairs 
of the $B_x$, $B_y$, and $B_z$ datacubes \citep[see][for details]{Liu_2012,Liu_2014}. 
The velocities were
further corrected by removing their components which are parallel to the
magnetic field \citep[see][]{Liu_2012,Liu_2014}. The helicity flux was
computed by integrating the so-called $G_{\theta}$ helicity flux density proxy 
\citep[see][]{Pariat_2005,Pariat_2006} over the area covered by the 
magnetograms. $G_{\theta}$ is given by

\begin{equation}
G_{\theta} ({\bf x}) = - \dfrac{B_{n}}{2 \pi} \int_{S'} \frac{d \theta({\bf r})}{dt} B'_{n} dS'  
\end{equation}
where $d \theta /dt$ is the relative rotation rate of two elementary
magnetic fluxes located at $\bf{x}$ and $\bf{x\arcmin}$ and $\bf{r}=\bf{x}-
\bf{x\arcmin}$. This rate does not depend on the choice of the direction that 
is used for the definition of $\theta$. The $G_{\theta}$
maps were derived by applying the fast Fourier Transform method of 
\citet{Liu_2013}. 

The accumulated changes in magnetic energy, $\Delta E$, and
helicity, $\Delta H$, were calculated by integrating the magnetic energy
and helicity fluxes over time. Following \citet{Thalmann_2021} the 
$\Delta E$ 
and $\Delta H$ time profiles have been constructed by using reference magnetic
energies and helicities that are equal to the corresponding average values of
these quantities deduced from the CB method over the first two hours of 
observations.

All magnetogram pixels were used for the calculation of the magnetic helicity 
and energy fluxes. For test purposes, in a few representative cases we took 
into account only those pixels that were used for the CB-method calculations 
(see Sect. 3.1), and found magnetic helicity and energy fluxes which were very 
close (differences of less than 1\%) to the ones obtained by the entire 
magnetograms' field of view.

\section{Long-term evolution of the magnetic helicity and energy of the ARs}

\subsection{Photospheric magnetic morphology}

Let us first discuss the evolution of the photospheric configurations of 
the two ARs (see Figs. 1 and 2, and also the associated movies).

AR11890 was classified as a $\beta\gamma\delta$ active region and produced 
several major flares and CMEs in early November 2103. One of its eruptive
events has been presented by \citet{xu2016} and \citet{Gupta_2021} while 
selected properties of the magnetic helicity injection rate in AR11890 have 
been discussed by \citet{Korsos_2020}.  

\begin{figure*}[!hb]
\centering
\includegraphics[width=16cm]{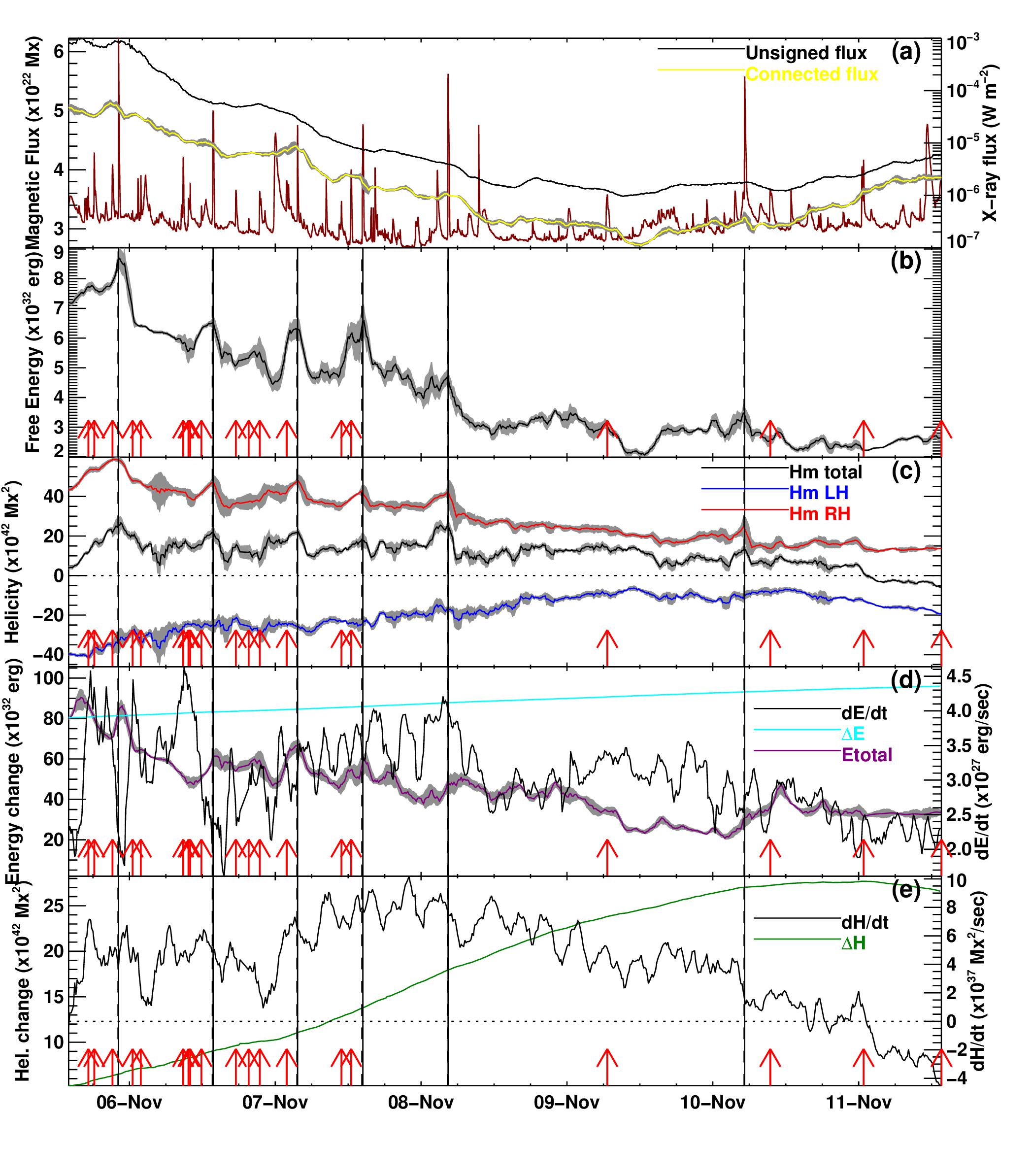}
\caption{Time profiles of magnetic properties of AR11890. (a) Total unsigned 
magnetic
flux, unsigned connected magnetic flux used in the CB-based method, and X-ray 
flux from the GOES 1-8 \AA\ channel (black, yellow, and maroon curves, respectively). (b) Free 
magnetic  energy. (c) Right-handed, left-handed, and net helicity (red, blue,
and black curves, respectively). (d) Magnetic energy injection rate, the 
corresponding accumulated energy, $\Delta E$, and the total energy from 
the CB method (black, cyan, and purple curves, respectively). (e) Helicity 
injection rate and the corresponding accumulated helicity, $\Delta H$, 
(black and green curves, respectively).
In this and subsequent figures, vertical straight lines indicate the start and
peak times of M- and X-class flares (all of them being eruptive) while 
arrows indicate the peak time of C-class flares (all of them being 
confined). The gray bands show the error bars.}
\end{figure*}

\begin{figure*}[!hb]
\centering
\includegraphics[width=16cm]{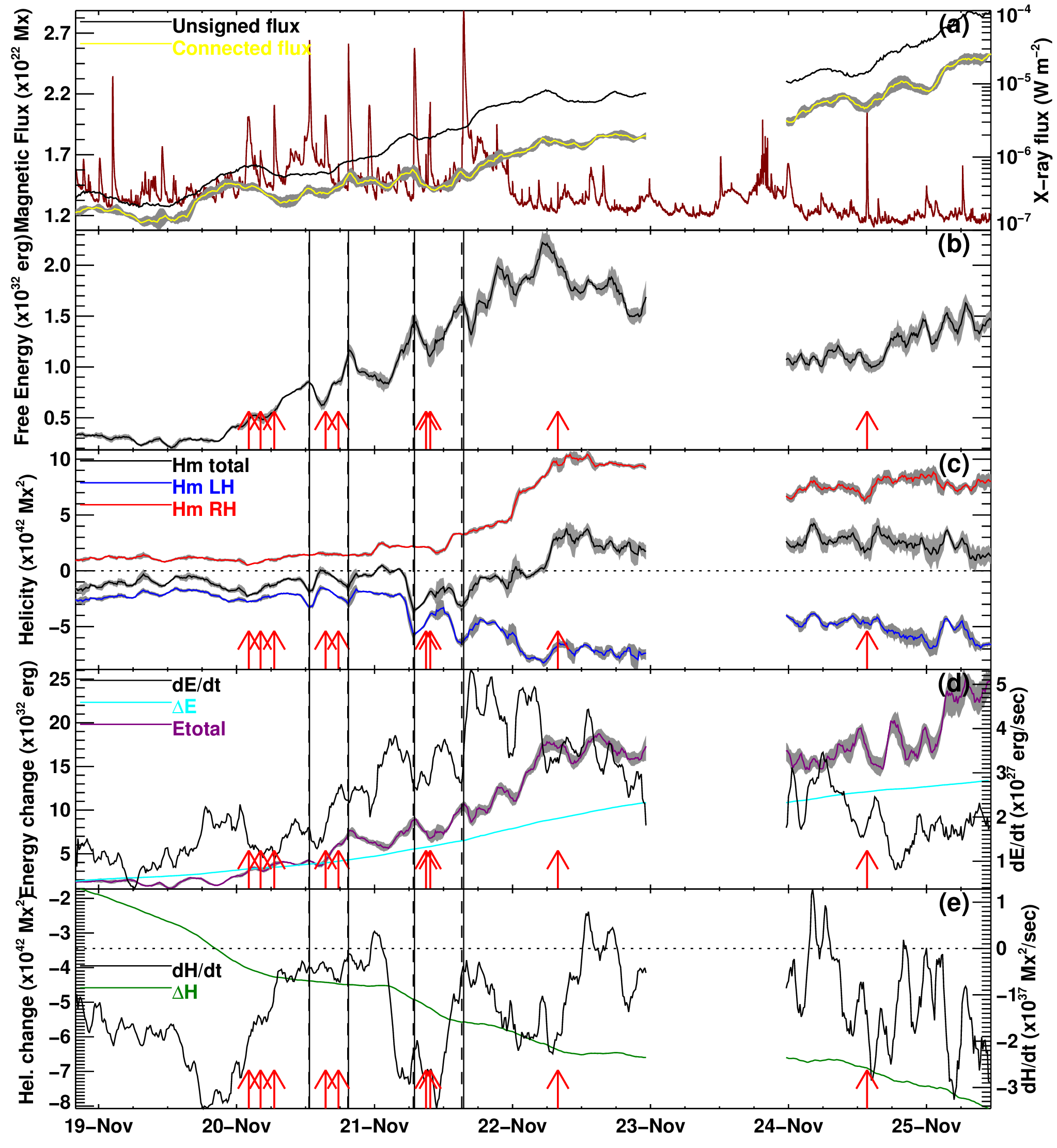}
\caption{Same as Fig. 3 but for AR11618.}
\end{figure*}

The evolution of the photospheric magnetic configuration of the AR is presented
in Fig. 1, and in the associated movie. At the beginning of observations 
(panel a) its major components are
a large fairly unperturbed preceding sunspot with positive polarity 
in the north-west part of the AR and a large bipolar following 
sunspot complex consisting of a massive negative polarity and a smaller 
elongated 
positive polarity patch (see the green arrow in panel b). Smaller patches of
positive and negative polarity, associated with smaller sunspots and pores, are
also located between these two large sunspots. 

At the first stages of the observations (panels a-h) the preceding positive
sunspot does not change much. However, magnetic flux decay is observed
in the eastern part of the AR between the elongated positive-polarity patch and the 
more massive negative polarity. As a result the massive eastern negative 
polarity gradually weakens (panels a-h) and attains an elongated shape.
At the same time due to shearing motions the eastern patch of positive polarity
gradually moves southwest of its initial location (compare the positions of
the arrows in panels b and f). From about November 9 09:30 UT onward, new 
positive magnetic flux emerges in the central part of the AR (see panels i-l)
while flux decay, albeit at a slower rate than before, continues 
to takes place in the eastern part of the AR. Furthermore, the large positive 
preceding sunspot appears to gradually develop a double umbra configuration 
(panels h-l).

The above trends are also illustrated in Fig. 3(a) where we show the evolution
of the total unsigned magnetic flux (that is, the algebraic sum of the 
positive magnetic flux and the absolute value of the negative magnetic flux) of
AR11890. The average flux decay rate in the interval defined from the
start of observations and the time when the flux attains its minimum value 
(November 9 09:36 UT) is $7.9 \times 10^{16}$ Mx s$^{-1}$. This value is almost 
an  order of magnitude larger than the largest cancellation rates measured by 
\citet{yardley_2018} in 20 small bipolar ARs. The corresponding decrease
in magnetic flux is  $2.6 \times 10^{22}$ Mx (in Fig. 3(a) 
compare the initial value of the unsigned flux with its minimum value) which 
amounts to about 40\% of the AR's initial
unsigned flux. This percentage lies on the high end of flux cancellation
percentages calculated by \citet{Green_2011}, \citet{Baker_2012} and 
\citet{yardley_2016,yardley_2018}. The subsequent flux emergence phase is
accompanied by cancellation and therefore the calculation of the average flux 
emergence rate from the time profile of the unsigned flux is not reliable.


Figure 2 and the associated movie show the evolution of the normal component 
of the 
photospheric magnetic field of AR 11618. It was an eruptive  AR which exhibited
multipolar photospheric magnetic configuration throughout the interval we 
studied. 

From about November 19 11:10 UT until the data gap (panels b-i) significant 
emergence of
both positive and negative polarity flux takes place primarily at the central
part of the AR. In panel (c) the major locations of flux emergence are marked 
with the red arrows; the left arrow shows the site of positive flux emergence
while the right one shows the site of negative flux emergence. As usually 
happens in the emergence phase \citep[e.g. see][]{Lidia_2015}
the magnetic polarities separate over time. Their separation brought them 
close to opposite polarity pre-existing fluxes. In both cases the proximity of 
opposite polarity fluxes gave rise to cancellation which resulted in the
gradual weakening of the fluxes that are enclosed in the yellow boxes of panel
(c). The combination of the above evolutionary trends resulted in the gradual
formation (see panels c-i) of two major sunspot groups. The first was 
located in the south-central part of the AR and consisted of three negative 
polarity sunspots with a common penumbra. The second was located in the 
central-eastern part of the AR and contained positive polarity sunspots with 
the exception of its small easternmost member which featured negative polarity. 
A new episode of flux emergence was captured after the data gap (panels i-l)
which is resulted in the enhancement of the magnetic flux at the 
central-eastern part of the AR.


The time profile of the unsigned magnetic flux in AR11618 appears in Fig. 4(a),
and reflects the major evolutionary trends discussed above. However, the fact
that before the data gap flux emergence is accompanied by cancellation (although
not as strong as flux emergence) makes the calculation of the corresponding 
average emergence rate from the time profile of the unsigned flux not reliable.
After the data gap, cancellation has been largely suppressed and the 
corresponding flux emergence rate is $4.5 \times 10^{16}$ Mx s$^{-1}$. This 
value is consistent with previous statistical studies
\citep[e.g.][]{otsuji2011statistical,Kutsenko_2019,Liokati_2022}.


\subsection{Diagnostics of evolution from magnetic helicity and energy}

In Fig. 3 and 4 we present the evolution of the free magnetic energy
(panels b) and helicity (panels c) of ARs 11890 and 11618, respectively. In all
free energy and helicity curves we show the 48-min averages of the 
actual curves in order to clearly assess the long-term evolution of both 
quantities. This can be described as the superposition of
slowly varying backgrounds featuring characteristic time scales of more than 
one day with shorter localized peaks which, in several cases, are associated 
with eruption-related changes.  

In more detail, the first phase of the evolution of $E_f$'s slowly varying 
component in AR11890 involves its decrease from the start of the observations
until about 09 November 2013 09:36 UT where it gets its minimum value. The
average rate of change referring to this period is $-1.4 \times
10^{27}$ erg s$^{-1}$. Then it
starts rising for about four hours and attains an extended plateau 
with an overall weak decreasing trend; 
this second phase lasts for about 25 hours. Subsequently a 
slow rise follows (rate of change of $3.2 \times 10^{26}$ erg s$^{-1}$) until 
the end of the observations. A comparison of the $E_f$ curve with the
black and yellow curves of Fig. 3(a) shows that these trends largely follow the 
large-scale temporal trends of both the total unsigned magnetic flux, $\Phi$, 
and the total magnetic flux that participates in the CB-method's connectivity 
matrix, $\Phi_{conn}$ 
\citep[hereafter referred to as connected flux; see][]{georgoulis2012magnetic}. 
During the flux decay phase the unsigned magnetic flux decreases and
so does the connected flux which results to the decrease of free energy while 
the opposite happens during the later flux emergence phase. The good 
correlation of the $E_f$ with $\Phi$ and $\Phi_{conn}$ is quantified by their 
linear (Pearson) and rank-order (Spearman) correlation coefficients which are 
0.92-0.88 and 0.80-0.74 for the $\Phi-E_f$ and $\Phi_{conn}-E_f$ pairs, 
respectively. Even higher ($\gtrsim 0.90$) correlation coefficients
are achieved if we consider the flux decay phase and the flux emergence
phase separately. Furthermore,
very similar correlation coefficients are derived if the $E_f$
timeseries is replaced by the total magnetic energy timeseries (see Fig. 3d).

The time profiles of AR11890's helicity (total, right-handed, and left-handed)
are presented in Fig. 3(c). Substantial values of both right-handed and 
left-handed helicity are present throughout the observations. The AR's net
helicity is right-handed (i.e. it has positive sign) from the start of the 
observations until about 11 November 2013 01:34 UT. Then it changes sign and
becomes left-handed (i.e. with a negative sign) until the end of the 
observations. This happens due to a
combination of two reasons. (1) During the flux decay both the 
right-handed and left-handed helicity decrease (in absolute values) 
because of the decrease of the 
unsigned and connected flux. However, the right-handed helicity decreases at a 
rate ($-6.3 \times 10^{37}$ Mx$^2$ s$^{-1}$) which is higher  
than the rate at which the left-handed helicity decreases ($5.5 \times 10^{37}$
Mx s$^{-1}$). (2) After $\Phi_{conn}$'s minimum and until the end of the 
observations, the slowly-varying component of the right-handed helicity does 
not show appreciable changes with time whereas the absolute value of the
left-handed helicity increases at a rate of $3.7 \times 10^{38}$ Mx$^2$ s$^{-1}$.
The net helicity change of sign explains its poor correlation with both the 
unsigned and connected fluxes (0.38 and 0.29 for the linear correlation
coefficient and 0.37 and 0.33 for the rank-order correlation coefficient). On 
the other hand, the right-handed and even more so the left-handed helicities
show much better correlation with the unsigned and connected flux (linear and
rank-order correlation coefficients in the range 0.68 to 0.93). Another 
simple explanation could be that the flux emerged from around midday on 
November 10 could be oppositely helical. This is also corroborated by Fig. 
3(e). 

The free energy and helicity evolution in AR11618 appear in Fig. 4(b) and 
(c), respectively and can be described as follows. The free energy from 
the start of 
the observations until about 19 November 2012 21:00 UT does not change 
appreciably. Then it 
increases until about 22 November 05:22 at a rate of $8.5 \times 10^{26}$
erg s$^{-1}$. This phase coincides with much of the flux emergence episode
before the data gap which led to the increase of both the unsigned and 
connected flux (see panel a). Subsequently, $E_f$ decreases from about 
noon UT on November 22 until the data gap.
This decrease does not match the corresponding evolution of the connected flux 
which shows a plateau. This behavior of $\Phi_{conn}$ is closer to the 
corresponding behavior of the total energy, $E_{tot}$ (see Fig. 4(d)). 
After the data gap both $E_f$ and $E_{tot}$ fluctuate for about 15.5 hours 
around values of 1.1 and $15.5 \times 10^{32}$ erg, respectively, and then 
increase
until the end of the observations. This behavior is broadly consistent with 
that of $\Phi_{conn}$ although the $E_f$ increase is milder. 
Overall,
the values of the linear and rank-order correlation coefficients are in the 
range 0.6-0.7 for the entire evolution of the $\Phi_{conn}-E_f$ pair. This 
correlation is lower than the one in AR11890 and results from the decrease
of the contribution of the free energy to the total magnetic energy budget after
the fourth major flare (this will become evident in the discussion of the
$E_f/E_{tot}$ time profile in Sect. 4.4). On the other hand, the correlation
coefficients of the $\Phi_{conn}-E_{tot}$ pair are higher than 0.9. This suggests
that in AR11618 the connected flux correlates better with the global magnetic 
field evolution \citep[see also][]{Tziotziou_2013}.


\begin{figure*}
\centering
\includegraphics[width=16cm]{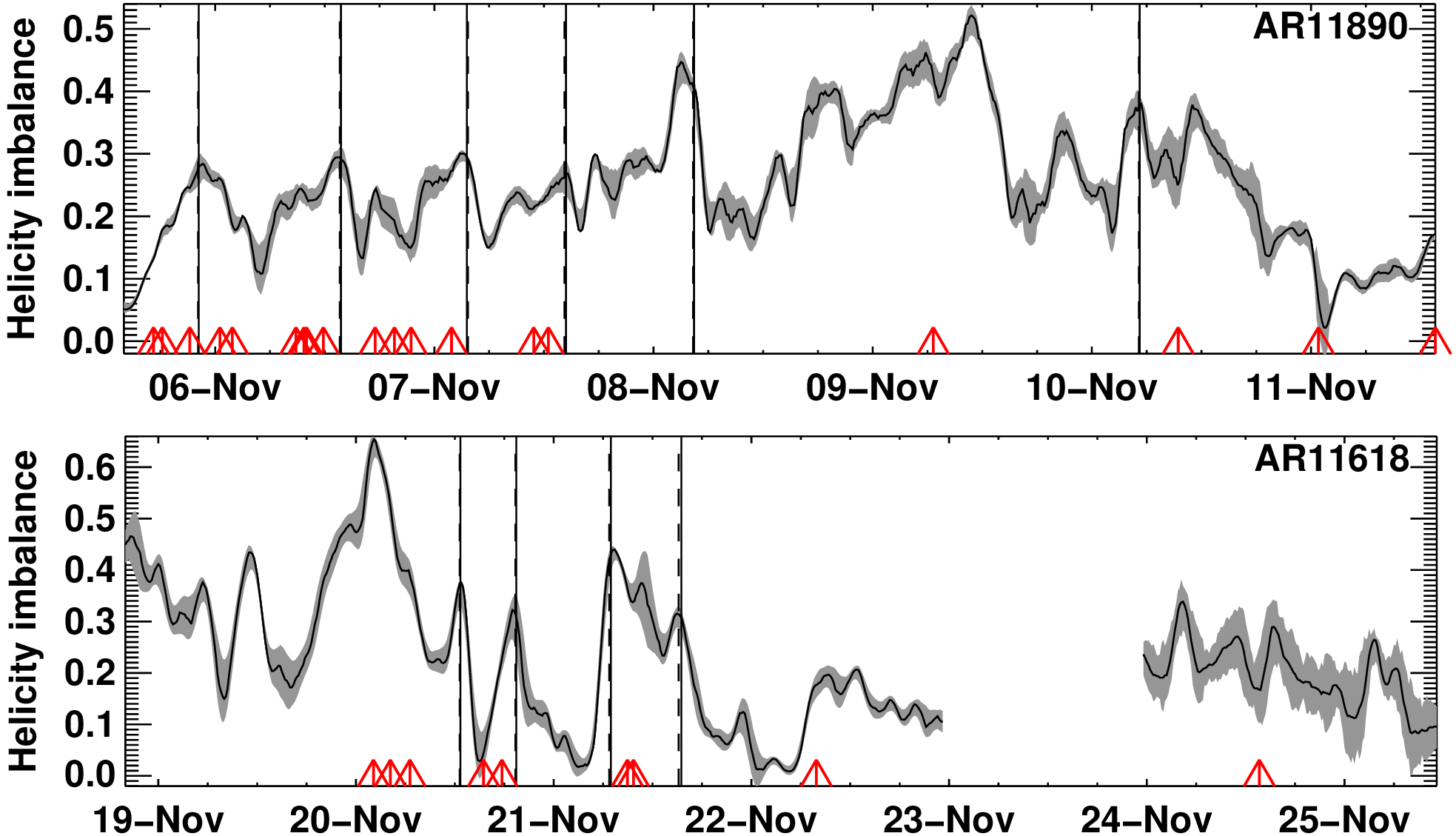}
\caption{Evolution of the helicity imbalance parameter, $h_{imb}$ (see text for
details) for AR11890 (top panel) and AR11618 (bottom panel).}
\end{figure*}

AR11618 shows predominantly (but weakly) left-handed net helicity from 
the start of the 
observations until about the time when the free energy gets its maximum value
(compare panels (b) and (c) of Fig. 4). Then the net helicity changes sign and 
becomes right-handed until the end of the observations. The time interval
of left-handed net helicity prior to the data gap is interrupted by three 
short 
excursions (centered at 20 November 14:46, 21 November 01:58, and 22 November 
00:46) where the net helicity becomes right-handed. Both the left-handed 
and right-handed
helicity do not show appreciable long-term variability from the start of the 
observations until about the end of November 20 although shorter-term changes 
associated 
with two of the four M-class flares are registered in this interval (see Sect.
5). Then both signed components of helicity increase by absolute value; the left-handed
helicity increases faster than the right-handed helicity in the interval until 
21 November 23:22 (increase rates of 4.3 and $3.7 \times 10^{37}$ Mx$^2$ s$^{-1}$,
respectively) and this combined with the initial predominance of the left sense
results in the left-handed sign of the net helicity. Subsequently, the 
situation is reversed 
in the interval until  22 November 13:22 (the corresponding rates of change are
$2.3 \times 10^{37}$ and $1.1 \times 10^{38}$ Mx$^2$ s$^{-1}$) resulting in the 
sign change of the net helicity at 22 November 04:22. Then both signed 
components of helicity do not show appreciable long-term changes until the data
gap. 

After the data gap, the right-handed helicity fluctuates around a value which 
is smaller than its value before the data gap ($\sim$8.0 versus $\sim$$9.4
\times 10^{42}$ Mx$^2$). The left-handed helicity emerges from the data gap with
a smaller absolute value than before (-4.2 versus $-7.5 \times 10^{42}$ Mx$^2$) 
which is also smaller in amplitude than the corresponding value of the 
right-handed helicity. The long-term amplitude increase of the 
left-handed helicity until the end of the observations (at a rate of $\sim$$-1.9
\times 10^{37}$ Mx$^2$ s$^{-1}$) is not adequate to 
balance the increase of the right-handed helicity and hence
the net helicity keeps its 
positive sign. As in AR11890, the correlation between the net helicity and
the connected flux is poorer (linear and rank-order correlation coefficients
of 0.60 and 0.65) than the resulting correlation of the left- or 
right-handed helicity with the connected flux (coefficients in the range 
0.70-0.88).

\begin{figure*}
\centering
\includegraphics[width=16cm]{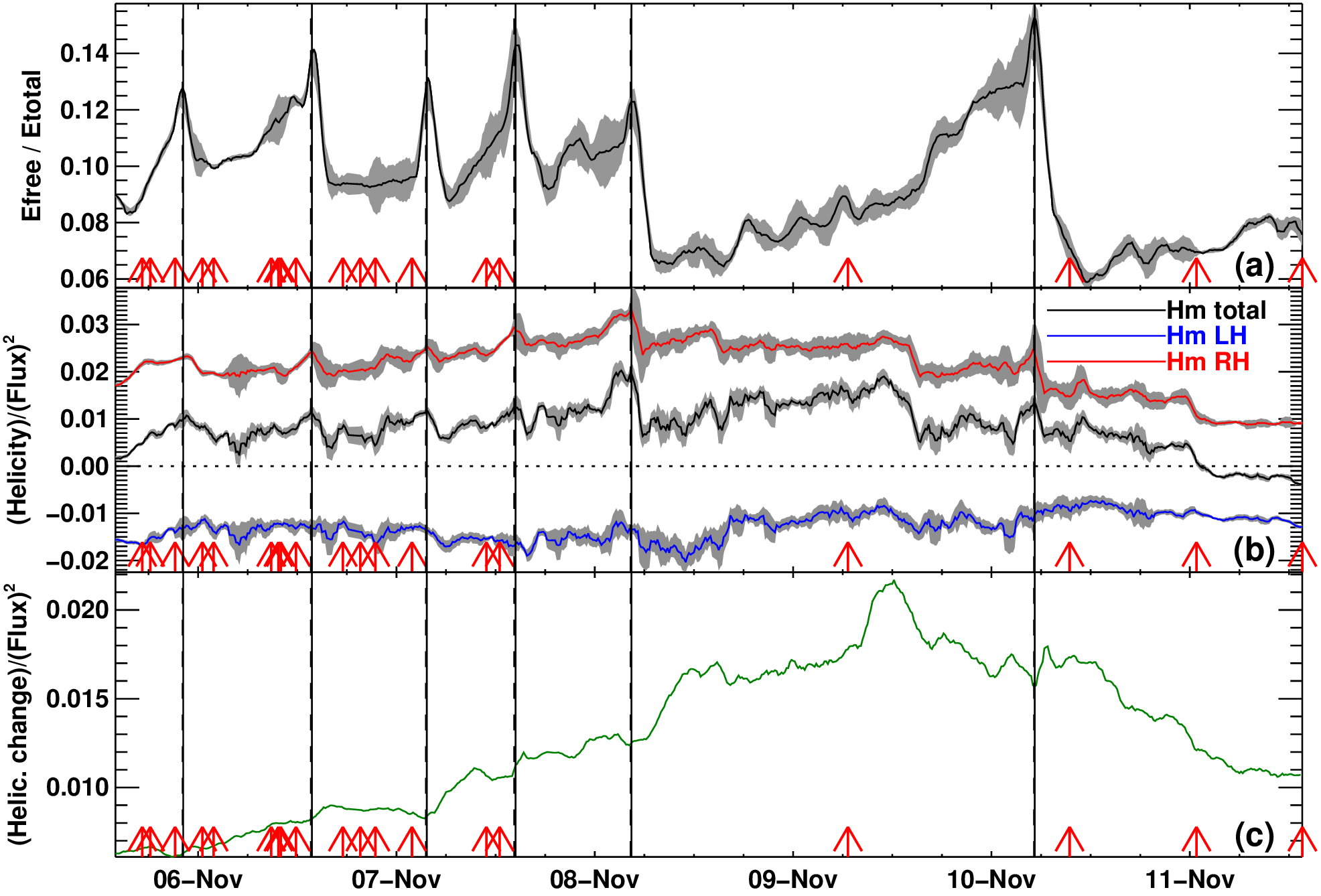}
\caption{Time profiles of normalized magnetic quantities for AR11890. (a)
Ratio of the free magnetic energy to the total magnetic energy. (b) Ratios
of the right-handed, left-handed, and net helicity to the connected magnetic 
flux squared (red, blue, and black curves, respectively). (c) Ratio of the
accumulated helicity (resulting from the helicity injection rate of the
flux-integration method) to the connected magnetic flux squared.}
\end{figure*}

\begin{figure*}
\centering
\includegraphics[width=16cm]{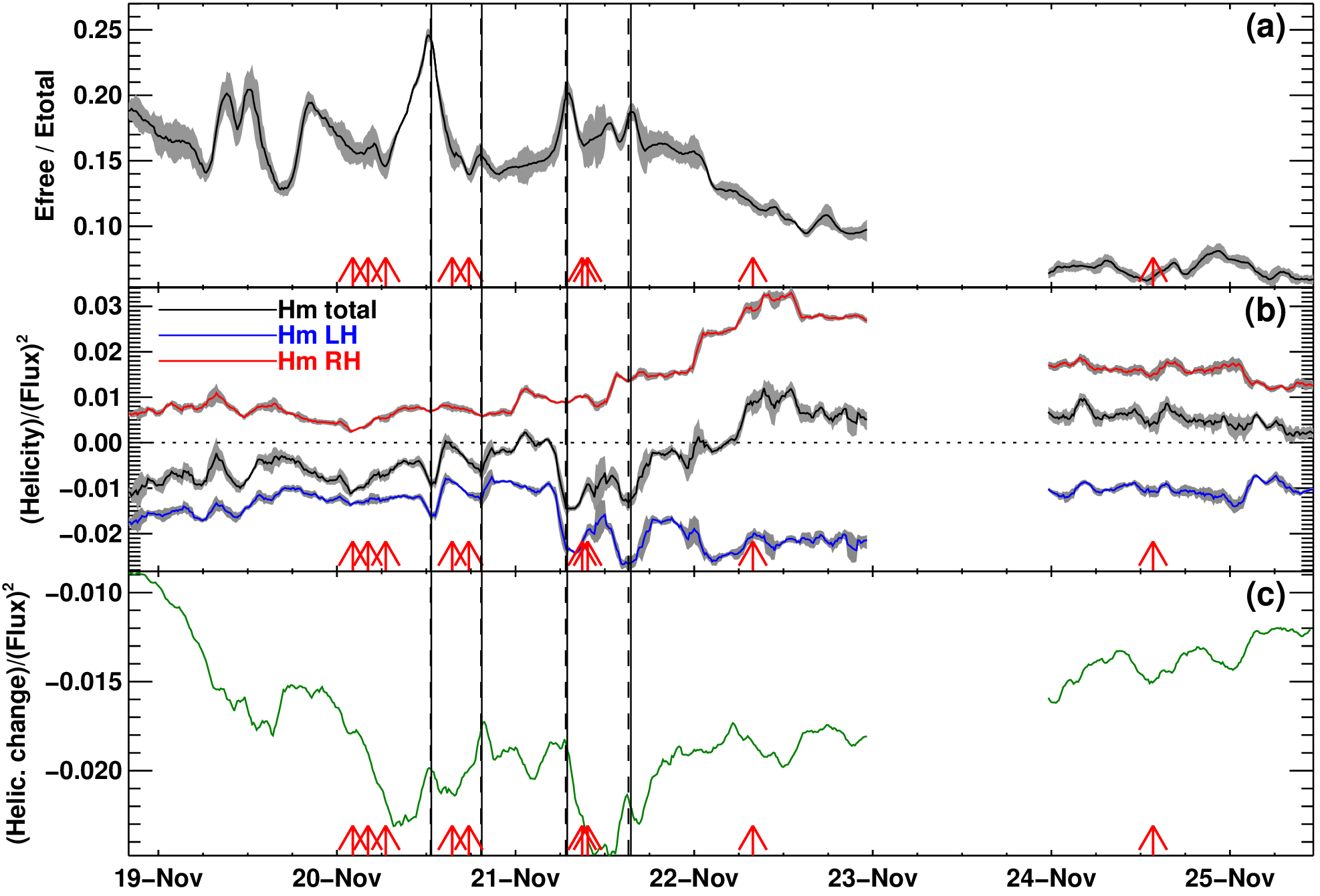}
\caption{Same as Fig. 6 but for AR11618.}
\end{figure*}

A worth mentioning property of the calculations presented in Figs. 3 and
4 is that both ARs contain substantial budgets of free magnetic energy during
much of the observations. In AR11890 the free energy is amply above 
the threshold of $4 \times 10^{31}$ erg for the occurence of major flares 
\citep[see][]{Tziotziou_2012}
throughout the observations and reaches a maximum value of $8.7 \times
10^{32}$ erg. The same is true for AR11618 after 19 November 2012 23:22; here
the free energy reaches maximum value of $2.2 \times 10^{32}$ erg. 
The signed terms of the net helicity also exhibit adequate
budgets \citep[that is, above the threshold of $2 \times 10^{42}$ Mx$^2$ for the occurrence of major flares;][]{Tziotziou_2012} throughout the observations. 
In AR11890 the positive and negative components of the helicity reach maximum 
values of $5.8 \times 10^{43}$ Mx$^2$ and $-4.1 \times 10^{43}$ Mx$^2$, 
respectively. The corresponding maximum values for AR11618 are $1.0 \times 
10^{43}$ Mx$^2$ and  $-8.2 \times 10^{42}$ Mx$^2$.

\subsection{Imbalance in the signed components of the helicity budgets}

A common feature of the helicity budgets of both
ARs is that throughout their evolution they contain comparable amounts of 
both right-handed and left-handed helicity. This can be quantified by 
introducing a helicity imbalance parameter, $h_{imb}$, 
\citep[see also][]{georgoulis2009}  as

\begin{equation}
h_{imb} = \frac{|H|}{H_+ + |H_-|}
\end{equation}
where $H$, $H_{+}$, and $H_{-}$ denote the net, right-handed, and left-handed 
helicity{, respectively. $h_{imb}$ can range from 0 (indicating perfect balance between the
positive and negative helicity) to 1 (indicating perfect dominance of a 
particular sense of helicity). The evolution of $h_{imb}$ for both ARs is
presented in Fig. 5. The plots indicate that during much of the
observations the value of $h_{imb}$ was below 0.5. The ARs acquire the minimum 
value of $h_{imb}$, which is zero, at the times when the net helicity changes
sign (see the discussion in Sect. 4.2). The temporal average of $h_{imb}$ was 
$0.25 \pm 0.09$ for AR11890 and $0.23 \pm 0.14$ for AR11618. 

Case studies 
\citep[e.g.][]{Pariat_2006,georgoulis2012magnetic,Tziotziou_2013,Vemareddy_2017,Vemareddy_2019,Thalmann_2019, Dhakal_2020}
indicate that most eruptive ARs feature a clear prevelance of one signed
helicity component over the other that does not change in the course of
observations. However, there are observations of ARs whose helicity sign 
changes during observations, but almost always these are non-eruptive ARs 
\citep[e.g.][]{Vemareddy_demoulin_2017,Vemareddy_2021,Vemareddy_2022}, 
although reports about one eruptive AR also exist 
\citep[see][]{georgoulis2012magnetic,Thalmann_2021}. The existence of a 
dominant sense of helicity in most eruptive
ARs throughout their observations is also supported either directly 
\citep[e.g.][]{Labonte_2007,georgoulis2009,Liokati_2022} or indirectly 
\citep[e.g.][]{Tziotziou_2012} by 
statistical studies. On the other hand, explicit reports about the relative 
contribution of the positive and negative helicity to the net
helicity budget of ARs are rather scarce. \citet{Tziotziou_2014} found that
the ratio between the $H_{+}$ and the $H_{-}$ terms of the net helicity
in the quiet Sun ranges from 0.32 to 2.31 with an average of 1.06. The temporal
averages of $h_{imb}$ that we found for our two ARs correspond to ratios of
$H_+/H_{-}$ of $1.68 \pm 0.46$ and $0.94 \pm 0.45$ for ARs 11890 and 11618,
respectively. Therefore, the helicity imbalance of our ARs is similar to that
of the quiet Sun, which is more exceptional, rather than nominal, for 
active regions.

\subsection{Diagnostics of evolution from magnetic helicity and energy
normalized parameters}

In order to put the free magnetic energy and helicity computations in the two 
active regions on the same footing, we calculated the following normalized
parameters: the ratio of the free magnetic energy to the total magnetic energy,
$E_f / E_{tot}$, as well as the magnetic-flux normalized net, right-handed, and 
left-handed helicity ($H / \Phi_{conn}^2$, $H_{+} / \Phi_{conn}^2$, and $H_{-} / \Phi_{conn}^2$, 
respectively). These quantities have often been used as proxies to quantify the
nonpotentiality of the magnetic field of ARs as well as their eruptive potential
\citep[e.g.][]{Pariat_2017,Thalmann_2019,Thalmann_2021,Gupta_2021}.
$E_f /E_{tot}$ quantifies the percentage of total magnetic energy that can be 
converted to other forms in flares and
CMEs. As a first approximation, the $H/ \Phi_{conn}^2$ parameter reflects the 
complexity of the magnetic
field structure while $H$ reflects both the structure and the flux budget;
this is because the helicity of an isolated flux tube with magnetic flux
$\Phi$ which is uniformly twisted with $N$ turns is simply $N\Phi^2$. 

The above normalized parameters for ARs 11890 and 11618 appear in Figs. 
6 and 7, respectively. We note that $E_f/E_{tot}$ acquires values in the range 
0.06-0.25 which are consistent with previous results 
\citep[e.g. see][]{Metcalf_1995,Guo_2008,Thalmann_2008,Malanushenko_2014,Aschwanden_2014,Gupta_2021}.
In both ARs the long-term evolution of the $E_f/E_{tot}$ curves is different 
from those of the $E_f$ curves. These different trends can be understood in 
terms of the varying contribution of the free energy to 
the budget of the total magnetic energy of the active regions.
In AR11890 there is no conspicuous decrease of $E_f/E_{tot}$ during much 
of the flux decay phase while in AR11618 there is
no conspicuous increase of $E_f/E_{tot}$ during the flux emergence phase before
the data gap. In AR11890 this happens because of the milder temporal decrease 
of 
$E_{tot}$ than $E_f$ (compare panels (b) and (d) of Fig. 3) while in AR11618 the
reason is the milder temporal increase of $E_{tot}$ than $E_f$ (compare panels 
(b) and 
(d) of Fig. 4). Furthermore, in AR11890 $E_f/E_{tot}$ gradually increases
in the interval starting after its downward jump that occurred after the fifth 
major flare and ending at the time of the sixth flare. Then it exhibits a 
sharp decrease for about 7.5 hours before resuming to show a mild increase 
until the end of the observations. In AR11618 $E_f/E_{tot}$ starts decreasing 
after the fourth major flare and until the end of the observations.

Going to the magnetic-flux normalized helicities (panels b of Figs. 6 and 7) we 
note that in both ARs $H_{\pm} / \Phi_{conn}^2$ possess absolute values in the range 
0.003-0.032 which are consistent with results presented in previous 
publications 
\citep[e.g.][]{Thalmann_2019,Thalmann_2021,Gupta_2021,Liokati_2022}.
In AR11890 there is a gradual increase of the $H_+/ \Phi_{conn}^2$ curve
from the start of the observations until the fifth major flare which is
followed by a gradual decrease until the end of the observations. On the
other hand the $H_{-}/ \Phi_{conn}^2$ curve is flatter than the $H_{-}$ curve (compare
panel b of Fig. 6 with panel c of Fig. 3). These
trends may reflect the gradual build-up of right-handed twist in the AR from
the start of the observations until the fifth major flare; after that time
it decreases with a rate which is faster than the decrease of the left-handed
twist (Sect. 4.2) resulting in the eventual sign change of the magnetic-flux
normalized net helicity. In AR11618 the magnetic-flux normalized helicities
show trends which are similar to those of their parent parameters (compare
panel c of Fig. 4 with panel b of Fig. 7). 

In agreement with the discussion in Sect. 4.2, in both ARs all major flares 
(which are all eruptive; see Sect. 2) are
associated with well-defined peaks of the helicity and free energy
normalized parameters (see Figs. 6 and 7). We will return to this 
significant finding in Sect. 5.

Several authors \citep[e.g.][]{Berger_2003,Moraitis_2014,Pariat_2017,Linan_2018,Linan_2020}
have decomposed the total helicity into a current-carrying
component and a volume-threading component (that is, the component of helicity
related to the field threading the boundary of the volume where the helicity
calculation is performed). However, the CB method cannot provide such 
decomposition and therefore we cannot test the popular parameter
(see Sect. 1 for references) defined by the ratio of current-carrying helicity 
to the total helicity. 


\subsection{Comparison of results from the connectivity-based and 
flux-integration methods}

Prior to a detailed discussion, the following notes are in order for the
results from the flux-integration method. First, a direct
comparison of the time profiles of the quantities provided by the two methods
is not straightforward because the flux-integration method yields the 
accumulated magnetic energy, $\Delta E$, and helicity, $\Delta H$, over certain
times, while the CB method yields their instantaneous 
budgets, including changes thereof. In essence, the CB method provides 
pseudo-timeseries (that is, timeseries in which each point is independent of 
the one before or the one after) contrary to the flux integration method. 
Second, the flux-integration method, in contrast to
the CB method, provides the total magnetic energy without separating into
free and potential components \citep[e.g.][]{Liu_2014}. 
Third, although all helicity
flux density proxies eventually should yield the same net helicity flux, its
distribution between its signed components may vary according to the flux
density proxy that was used \citep[][]{Pariat_2005,Pariat_2006,dalmasse2014}. 
These authors,
as well as \citet{dalmasse2018}, have pointed out that the $G_{\theta}$ proxy 
may yield fake helicity flux density polarities.
This is why studies, with this one among them, present only net rates or 
integrated helicities from the flux-integration method. Fourth, the
uncertainties of the magnetic energy injection rate, $dE/dt$, and the
helicity injection rate, $dH/dt$, were calculated for selected representative
values of these quantities (that is, at a nonuniform sampling)  by using the
Monte Carlo experiment approach as in \citet{Liokati_2022}
\citep[see also][]{Liu_2012,Liu_2014}. They never exceeded 20\%. Furthermore,
the corresponding uncertainties in the accumulated quantities, $\Delta E$ or
$\Delta H$, were about two orders of magnitude smaller than the pertinent
accumulated quantity. In the pertinent curves of Figs 3(d-e), 4(d-e),
6(c), and 7(c) we do not mark the uncertainties of the flux integration results
because of their small values and nonuniform sampling.

The time profiles of the magnetic energy and helicity injection rates from the 
flux-integration method as well as the corresponding accumulated quanties are 
presented in panels (d) and (e), respectively, of Figs 3 (for AR11890) and 4 
(for AR11618). 
In panels (c) of Figs. 6 and 7 we also show the magnetic-flux normalized
profiles of the accumulated $\Delta H$ from the flux-integration method in ARs 
11890 and 11618, respectively.

In both ARs, both $dE/dt$ and $dH/dt$ exhibit significant short-term 
fluctuations lasting up to 10 hours which are superposed on their longer-term 
evolution. Most of these fluctuations do not seem directly relevant to any of 
the major flares of the ARs. Therefore, in most cases, our results are not 
necessarily consistent with previous reports of occasional good 
correlation between helicity injection and soft X-ray activity 
\citep[e.g.][]{Maeshiro_2005}. 
However, we note that \citet{Korsos_2020} have advocated for
periodicities of 8 and 28 hours prior to three X-class flares of AR11890. 


\begin{figure*}
\centering
\includegraphics[width=18cm]{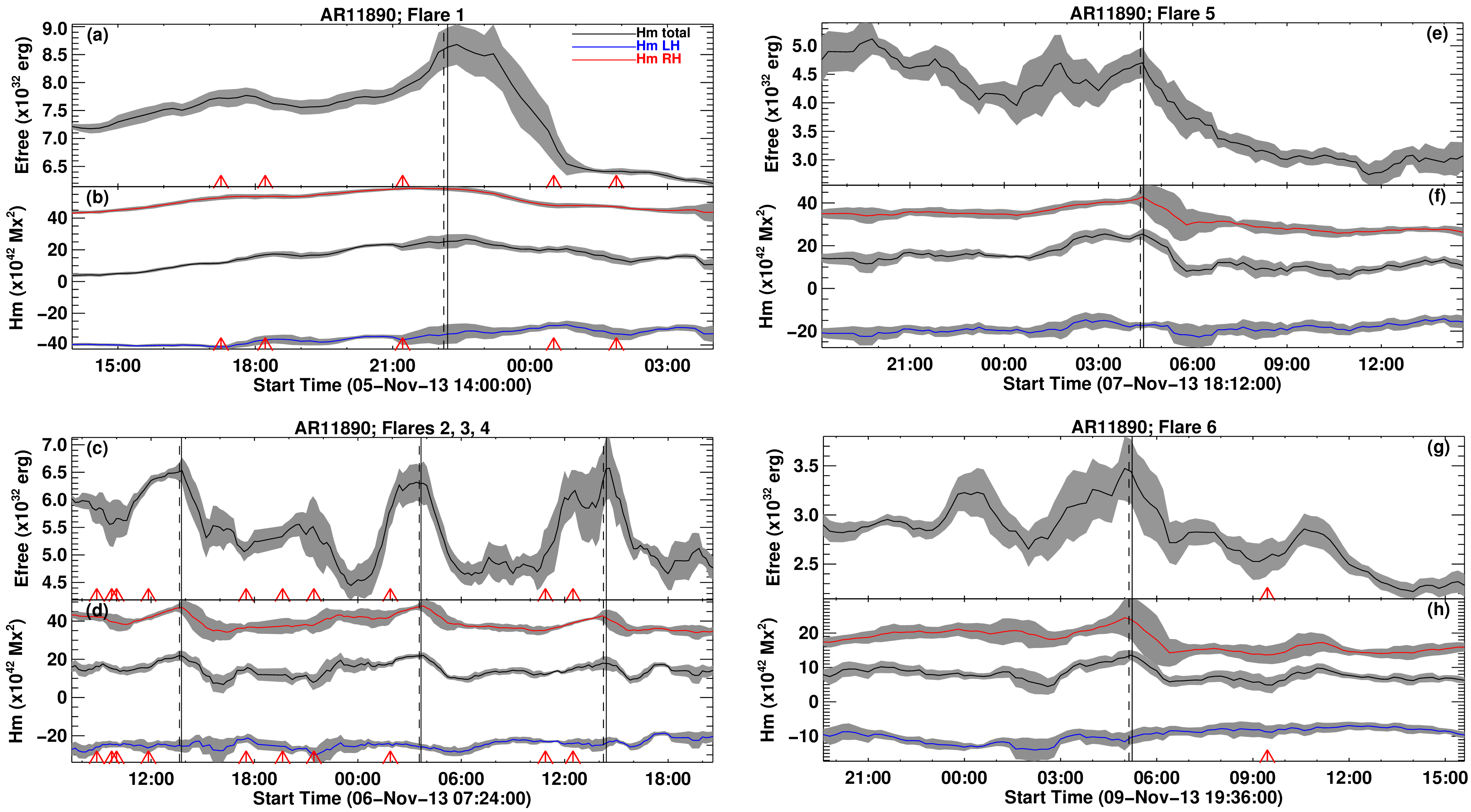}
\caption{Stack of plots showing the evolution of free magnetic energy (top plot) and of the net, right- and left-handed helicity budgets (bottom plot) for each of the six major eruptive flares in AR11890 a few hours before and after the events. The first (X3.3) flare is shown in plots (a, b), flares 2, 3 and 4 (M3.8, M2.3 and M2.4, respectively) are shown in plots (c, d), the fifth flare (X1.1) is shown in plots (e, f), while the sixth flare (X1.1) is shown in plots (g, h).}
\end{figure*}

\begin{figure}
\centering
\includegraphics[width=9cm]{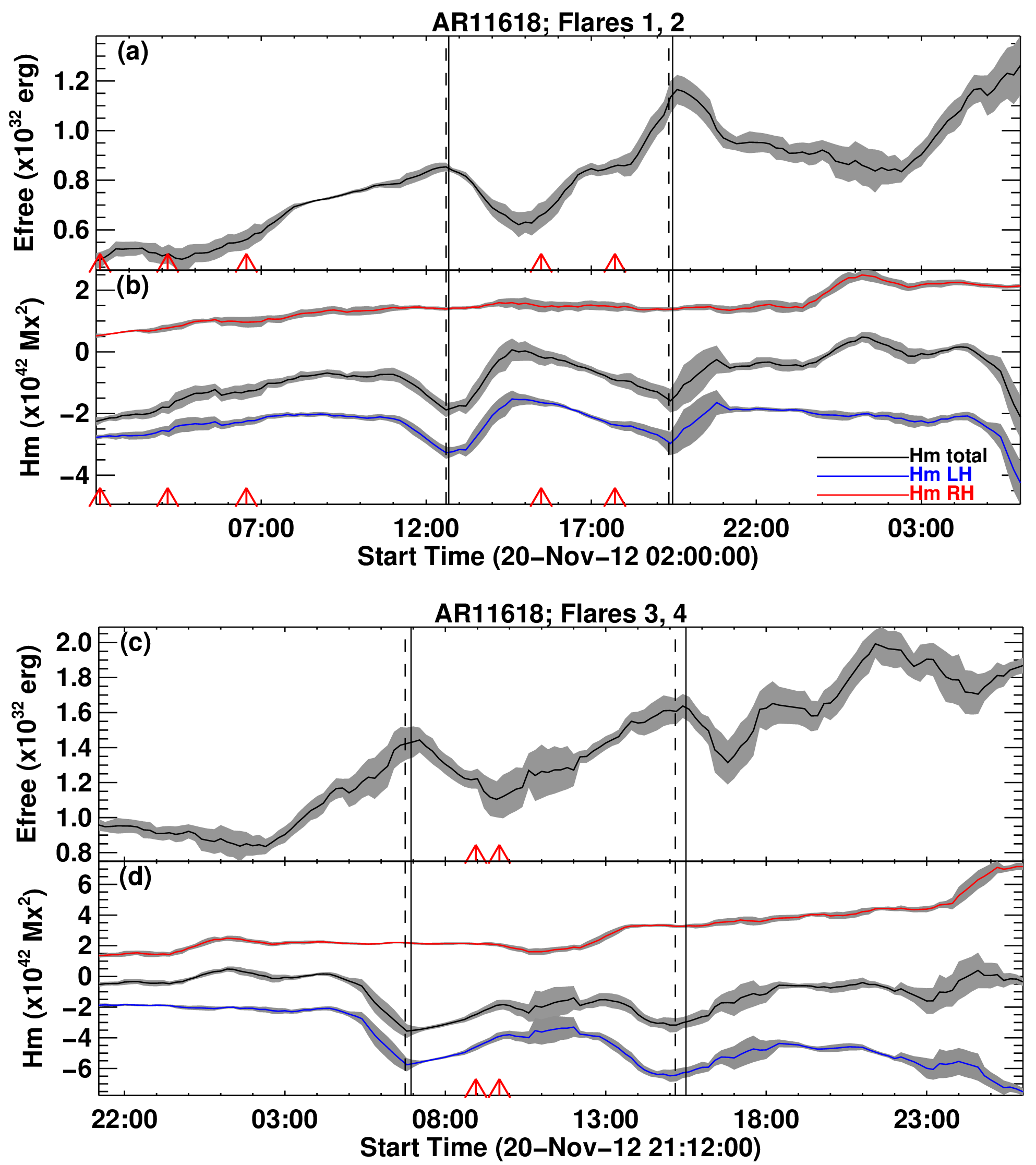}
\caption{Stack of plots showing the evolution of free magnetic energy (top plot) and of the net, right- and left-handed helicity budgets (bottom plot) for each of the four major eruptive flares in AR11618 a few hours before and after the events. The first and second flares (M1.7 and M1.6, respectively) are  shown in plots (a, b) while the last two flares (M1.4 and M3.5, respectively) are shown in plots (c, d).}
\end{figure}

In AR11890, and in agreement with the CB-method results, $dH/dt$ is 
positive for much of the observations. It changes sign for the first
time some 10 hours prior to the sign reversal of the instantaneous helicity.
This 10-hour interval is characterized by very small $dH/dt$ values and multiple
changes of its sign.
Subsequently, the $dH/dt$ sign stays negative until the end of the
period we studied, in agreement with both the sign and the overall weak
decreasing trend of the instantaneous helicity (compare panels (c) and (e) of
Fig. 3).
Contrary to the $E_{tot}$ time profile from the CB method, the $\Delta E$ curve 
steadily increases with time because it is constructed by the accumulation of 
positive values (see Fig. 3(d)). The $\Delta H$ curve shows a similar behavior 
until around the time of the sign reversal of the 
helicity. Then it shows a plateau and subsequently declines until the end of 
the observations because negative 
values start accumulating into the helicity budget. However, these negative 
budgets are not sufficient to change its positive sign. At the end of the
observations $\Delta H$ has acquired a value of $2.6 \times 10^{43}$ Mx$^2$
which is a factor of 1.8 higher than the corresponding instantaneous budget
of $H_+$. 

The $\Delta H / \Phi_{conn}^2$ curve of Fig. 6(c) exhibits 
richer structure than the $\Delta H$ curve of Fig. 3(e) which reflects the 
competition 
between the increasing $\Delta H$ evolution and the more complicated evolution
of the $\Phi_{conn}$ time series 
We also note that the 
$H_+ / \Phi_{conn}^2$ and $\Delta H / \Phi_{conn}^2$ values at the end of the 
observations are similar (differences of 17\%).


In AR11618 $dE/dt$ increases at least until about the fourth major flare
(see Fig. 4(d)). 
The evolution of the corresponding accumulated energy, $\Delta E$ 
is qualitatively similar to that of the instantaneous total magnetic energy, 
$E_{tot}$, from the CB method.
This could not happen if $E_{tot}$ were decreasing with time (see the relevant discussion about AR11890). At the end of the observations, $E_{tot}$ has acquired a value of $2.4 \times 10^{33}$ erg while $\Delta E$ a value of $1.3 \times 10^{33}$ erg. 



The situation is more complicated regarding the comparison of the helicities 
from the two methods in AR11618.
$dH/dt$ maintains a negative sign throughout much of the observations with only
three relatively short positive excursions. These intervals 
coincide with intervals where the instantaneous net helicity is positive,
however after the fourth major flare the 
instantaneous net helicity is positive for much more time (compare panels
(c) and (e) of Fig. 4). This discrepancy would be partly reconciled if we 
could assume that strong injection of positive helicity took place during the 
data gap which gradually decreased and
eventually changed sign after the data gap. This speculation is consistent
with the declining trend appearing in the instantaneous net helicity after the
data gap. Finally, Fig. 7(c) indicates that, in agreement with the results from
the CB method,  the $\Delta H/ \Phi_{conn}^2$ exhibits enhanced left-handed 
budgets in an extended interval that includes the AR's four major flares.

\section{Helicity and energy budgets of major eruptive flares}


The start and peak times of the eruptive flares that occurred in the two
ARs (see Sect. 2) are marked with dashed and solid vertical lines in
Figs. 3 and 4 (due to the scales of these figures the dashed lines 
practically coincide with the solid lines) and are also given in 
Tables 2 and
3. 

In AR11890 all but the last major flare occur during the flux decay 
phase of the
AR during the long-term decrease of both the free energy and the signed 
components of the helicity. On the other hand, in AR11618 they occur during its
first flux emergence phase when both the free energy and the (absolute) negative
helicity increase (see also the discussion in Sect. 4.2). 

In both ARs the major eruptive flares have a significant imprint in
the time profiles of the magnetic helicity and free energy budgets. Figs. 3 and
4 indicate that several well-defined peaks of both the helicity and free energy 
time profiles are associated with the occurrence of the eruptive flares. This
shows better in Figs. 8 and 9 where we present the evolution of the
helicity and free energy a few hours before and after the major flares. 
In most cases both quantities show localized peaks that occur around the 
impulsive
phase of the flares (that is, the interval between the start and peak times of
the flares). Small temporal offsets are visible in a few cases (compare the 
positions of the vertical lines with the times of the $E_f$ and $H$ peaks
in Figs 8-9) 
With the exception of the $H$ peak of AR11890's flare 1 (see Fig. 8(b)) these 
temporal offsets are of the order 12 minutes and, therefore, barely 
resolved given the cadence of the magnetograms.  In flare 1 of AR11890 the 
net $H$
curve peaks about 30 minutes after the flare maximum but the prevailing
left-handed helicity exhibits a broad peak in an interval that contains the
impulsive phase of the flare. 


The peaks of the net helicity associated with the occurrence of the major
flares are accompanied by peaks of the corresponding prevailing
component of the helicity (positive for AR11890 and negative for AR11618). We 
note that here, and in what follows, the word ``prevailing'' (``nonprevailing'')
is used to describe the component of helicity (be it positive or negative) with
the largest (smallest) absolute value at a given time. In practically all cases
this behavior does not reflect on the nonprevailing component of the 
helicity 
which exhibits small, apparently unrelated temporal changes around the 
flare times. 
Directly related to the above discussion is the fact that all major eruptive
flares occur at times of local helicity imbalance enhancements (in Fig. 5
compare the morphology of the $h_{imb}$ curves with the location of the vertical
lines). This result however, should not be overinterpreted because Fig. 5 indicates that
there are extensive intervals (e.g. several hours between the fifth and sixth 
major flares in AR11890 and several hours before the first major flare in 
AR11618) in which $h_{imb}$ attains large values unrelated to any 
eruptive activity.
Point taken, there is not a single case of occurrence of a major flare when
$h_{imb}$ exhibits a local minimum.

It is interesting that in both ARs the time profiles of the total magnetic 
energy (panels (d) of Figs. 3 and 4) also exhibit, in most cases, local peaks 
possibly associated with the occurrence of the eruptive flares. The only exception are
flares 5 and 6 of AR11890 and probably flare 1 of AR11618. This result 
indicates that in most cases the free and total magnetic energy may
vary in phase 
around the times of eruptions. 

In both ARs the occurrence of the confined C-class flares (their peak times are
marked by the red arrows in the figures) is usually not associated with any 
prominent signature in the evolution of the free magnetic energy and helicity 
budgets. However, the following exceptions have been registered. A clear free 
energy enhancement accompanied by a helicity plateau are associated with the 
first two C-class flares that occurred about 4-5 hours prior to the first 
X-class flare in AR11890 (see Figs 3 and 8a,b). 
There is a broad free energy peak between the second and third major
flares in AR11890 (see Fig. 3 and 8c) which is probably associated with the
occurrence of three C-class flares. At the same time the helicity curves do not
show significant changes in agreement with the well-known fact that confined
flares do not remove any helicity. Furthermore, in AR11618 the first three
C-class flares occur around the time of the first free energy local peak or 
shortly thereafter (see Fig. 4(b)). This free energy peak is also related to 
an (absolute value) helicity peak (see also the corresponding enhanced values 
of $dH/dt$ in Fig. 4(e)). It is possible that the overlying magnetic field 
could inhibit eruptions in these cases.

Concerning the free magnetic energy and helicity budget associated with the 
major flares
we first note that, with the possible exception of flare 2 of AR11618, all 
major events occur at times when the ARs possess adequate net magnetic 
helicity and
energy budgets which exceed thresholds defined in previous publications 
\citep[see][and also the discussion in Sect. 4.2]{Tziotziou_2012,Liokati_2022}.
Flare 2 of AR11618 occurs when the free energy budget of the AR exceeds
the $4 \times 10^{31}$ erg threshold proposed by \citet{Tziotziou_2012}, but
the corresponding helicity budget of the AR is between the thresholds proposed 
by the above authors ($2 \times 10^{42}$ Mx$^2$ and $9 \times 10^{41}$ Mx$^2$,
respectively). This said, the individual budget of the prevailing sense 
of helicity (negative) was $-3.1 \times 10^{42}$ Mx$^2$ and exceeded the 
threshold at the time of the flare.

\begin{figure}
\centering
\includegraphics[width=9cm]{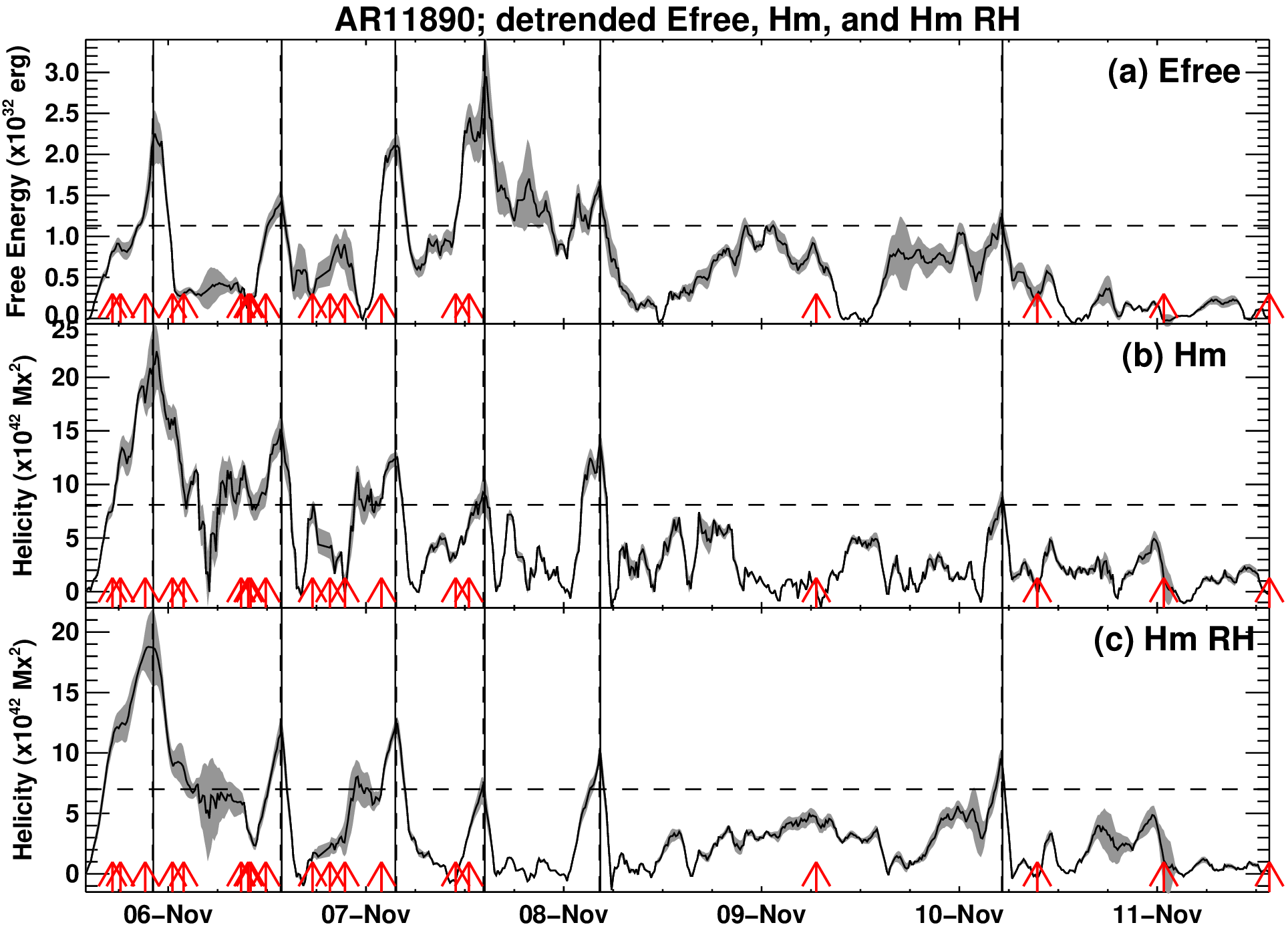}
\caption{Detrended time series (see text for details) of the free magnetic 
energy (panel a), the net helicity (panel b), and
the prevailing signed component of helicity (right-handed, panel c) for AR11890. 
The dashed lines denote the $2 \sigma$ level of the quantities above 
the (removed) background.}
\end{figure}

\begin{figure}
\centering
\includegraphics[width=9cm]{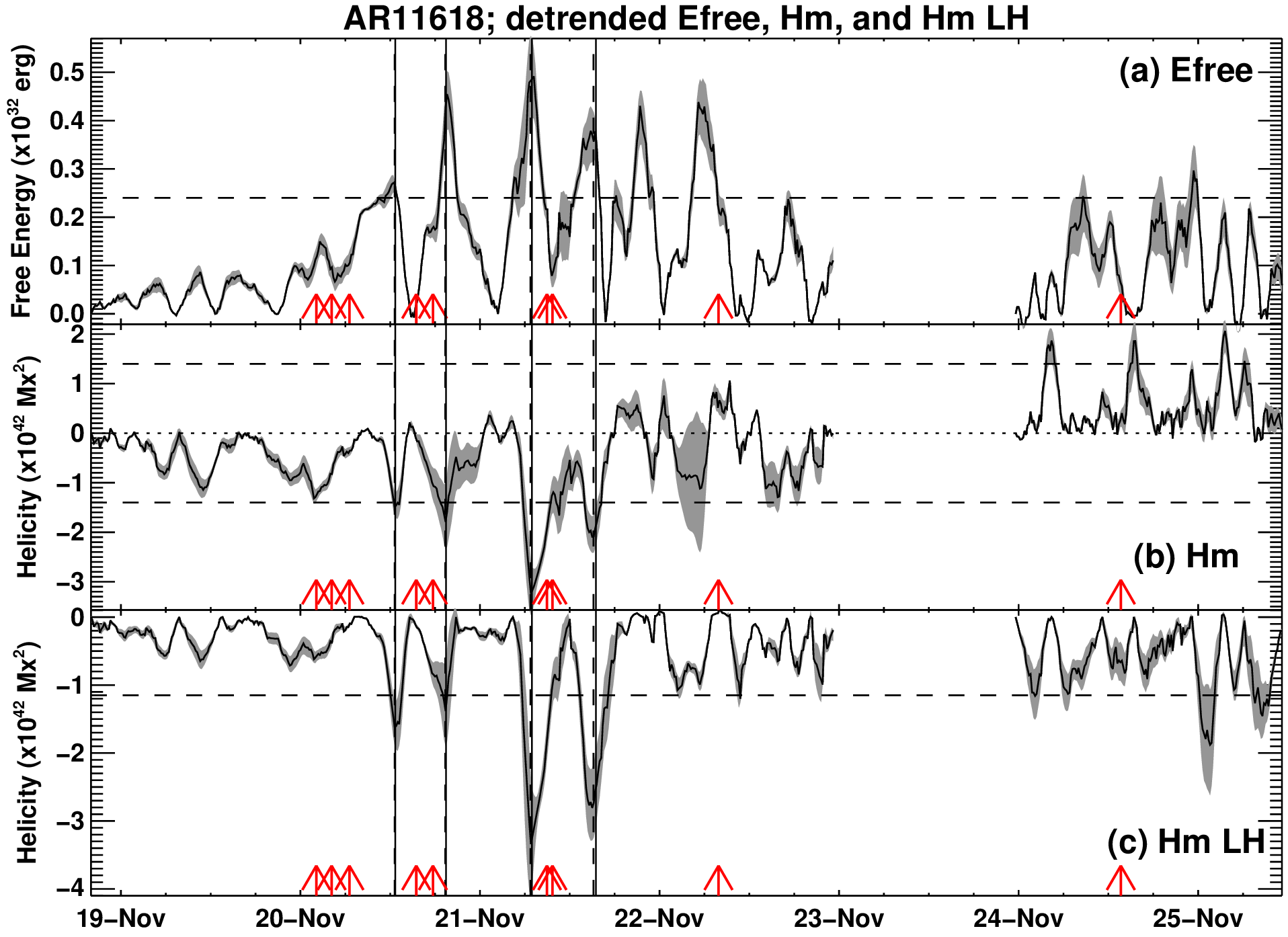}
\caption{Same as Fig. 10  but for AR11618, with the exception that in panel (c)
the left-handed helicity is displayed.  The dashed lines denote the $\pm 2 
\sigma$ level of the quantities and the dotted line the zero helicity.}
\end{figure}

\begin{table*}
\begin{center}
\caption{Eruptive flares in AR11890}
\begin{tabular}{ccccccc}
\toprule
Event  & Flare start and maximum  & Flare class  & $\Delta E_f$ & $\Delta H$ & $\Delta E_f$ & $\Delta H$  \\
Number &      (UT)                &              & ($\times 10^{32}$ erg) & ($\times 10^{42}$ Mx$^2$) & (\%) & (\%) \\
1      & 2013 Nov. 05 22:07 22:12 & X3.3 & 2.0  & 20.6 & 20 & 25 \\
2      & 2013 Nov. 06 13:39 13:46 & M3.8 & 1.1  & 14.9 & 33 & 62  \\
3      & 2013 Nov. 07 03:34 03:40 & M2.3 & 1.5  & 11.9 & 33 & 50 \\
4      & 2013 Nov. 07 14:15 14:25 & M2.4 & 1.8  & 8.8 & 35 & 42 \\
5      & 2013 Nov. 08 04:20 04:26 & X1.1 & 1.5  & 14.8 & 46 & 66 \\
6      & 2013 Nov. 10 05:14 05:18 & X1.1 & 0.8  & 7.6  & 61 & 48 \\
\bottomrule
\end{tabular}
\end{center}
\end{table*}

\begin{table*}
\begin{center}
\caption{Eruptive flares in AR11618}
\begin{tabular}{ccccccc}
\toprule
Event  & Flare start and maximum  & Flare class  & $\Delta E_f$ & $\Delta H$ & $\Delta E_f $ & $\Delta H$ \\
Number &      (UT)                &              & ($\times 10^{32}$ erg) & ($\times 10^{42}$ Mx$^2$) & (\%) & (\%) \\
1      & 2012 Nov. 20 12:36 12:41 &  M1.7 & 0.28  & -1.65  & 37 & 98 \\
2      & 2012 Nov. 20 19:21 19:28 &  M1.6 & 0.45  & -1.28  & 10 & 82 \\
3      & 2012 Nov. 21 06:45 06:56 &  M1.4 & 0.41  & -2.70  & 20 & 36 \\
4      & 2012 Nov. 21 15:10 15:38 &  M3.5 & 0.38  & -2.56  & 16 & 83 \\
\bottomrule
\end{tabular}
\end{center}
\end{table*}


To better reveal the local peaks of free energy, net helicity, and
prevailing signed component of helicity budgets (that is, right-handed for
AR11890 and left-handed for AR11618)
associated with the eruptive flares, we separated them from the long-term 
slowly varying background evolution of the respective time series.
The background
subtraction was done by fitting spline curves to the free energy and helicity
time series that appear in Figs 3 and 4. We followed such a procedure because
it was not possible to fit the slowly-varying background by using a single 
polynomial throughout a given time series. The resulting detrended curves
appear in Figs. 10 for AR11890 and 11 for AR11618. In order to check the 
reliability of our background subtraction scheme we compared the values of the 
peaks in the detrended time series with those resulting from the following
procedure. In the time series of Figs. 3 and 4 we found the inflection
points just before and just after each major eruption-related local peak, 
calculated the 
corresponding average $E_f$, $H$ and $H_{\pm}$ value which was then subtracted from the 
pertinent local peak. The two methods yielded similar values for the local 
peaks (differences of up to about 10\%).

In the detrended time series, the eruption-related changes were calculated as 
the difference between the pertinent local peak and the value at the inflection
point just after the local peak. The results for ARs 11890 and 11618 appear
in the fourth and fifth columns of Tables 2 and 3, respectively. These free 
energies are broadly consistent 
with previous results from magnetic field extrapolations
\citep[e.g. see][]{Emslie_2005,Emslie_2012,Sun_2012,Tziotziou_2013,Aschwanden_2014,Thalmann_2015} while the helicities are broadly consistent with reported
helicities of magnetic clouds 
\citep[e.g. see][]{Lepping_1990,Lepping_2006,DeVore_2000,Demoulin_2002,Lynch_2003,georgoulis2009,Patsourakos_2016}. 

Figs. 10 and 11 indicate that all eruptive events occurred at times when
the free magnetic energy, net helicity, and prevailing signed component of helicity local peaks exceed the 2$\sigma$ 
level ($\sigma$ denotes the standard deviation) of the pertinent time series.
The 2$\sigma$ levels are marked with horizontal dashed lines in Figs. 10 and 
11. It is interesting that in cases where one of $E_f$ or $H$, but not both, 
exceed the 2$\sigma$ level, no eruptive activity is registered. The
same conclusion holds if we replace $H$ with the prevailing signed component
of helicity.

By fitting gaussians to the eruption-related components of the 
detrended timeseries of Figs. 10 and 11, we found that their full width at 
half maximum (FWHM) are in the range 3.7-8.1 hours. No essential differences
were found between $E_f$, $H$, and prevailing signed component of
$H$.

We also note that the time profiles of the normalized parameters 
related to the free magnetic energy, the net helicity and the prevailing signed
component of helicity (see Figs. 6 and 7, and the discussion in Sect. 4.4) also
show well defined local peaks associated with the major eruptive flares. 
In the sixth and seventh columns of Tables 2 and 3 we give the 
corresponding percentages of $E_f$ and $H$ losses (in
their normalized parameters) associated with the major eruptive flares of 
AR11890 and 11618, respectively.

\section{Summary and conclusions}

Using the CB method by \citet{georgoulis2012magnetic}, we studied the free 
magnetic energy and helicity in two differently evolving eruptive ARs, AR11890 
and 11618. Using this calculation we were able to identify all 
major patterns of photospheric magnetic field evolution (see Sect. 1). However,
it is clear that intense flux decay dominated the evolution of AR11890 
for more than half of the observations. Flux decay was later paired with
flux 
emergence until the end of the observations. AR11890 was the site of six major
eruptive flares (three X-class and three M-class); all but the last one occurred
during the flux decay phase. On the other hand the evolution of AR11618
was dominated primarily by flux emergence. This AR was the site of four eruptive
M-class flares. 

In both ARs, the evolution of the free magnetic energy and helicity can be
understood in terms of the superposition of a slowly varying component (with
characteristic time scales of more than 24 hours) and localized peaks, some of
which associated with eruptive flares (the characteristic time scales of 
the apparently eruption-related energy and helicity changes are 3.7-8.1 hours). In both ARs the evolution of the total 
magnetic energy is largely consistent with the evolution of the connected 
magnetic flux. 
The same is true for the evolution of the free
energy in AR11890 but not in AR11618; in the latter case the correlation 
worsens.

The long-term evolution of both signed components of the helicity is also 
consistent with the evolution of the connected flux. However, this is not the
case for the net helicity because in both ARs it changes sign during the
observations. It is believed 
\citep[see][]{Vemareddy_demoulin_2017,Vemareddy_2021,Vemareddy_2022}
that ARs featuring
successive injection of opposite helicity do not produce CMEs. Our study 
shows two remarkable counter-examples; in addition to the accumulation
of substantial free energy budgets, our ARs also accumulate substantial 
amounts of net helicity to enter them in eruptive territory despite the fact 
that relatively shortly thereafter (19 hours in AR11890 and 14 hours in 
AR11618) their helicity changes sign. This is one of the important findings of 
this work because (1) such reports are rare, and (2) the ARs involved featured 
different patterns of photospheric magnetic field evolution.

Our study was not only able to detect the change of the net helicity sign during
observations but also (thanks to the properties of the CB method) to unravel, 
at any given time, the relative contributions of the signed components of 
helicity to the net helicity budget. The helicity imbalance parameter that we 
used in Sect. 4.3 indicates that throughout the evolution of the ARs the 
prevelance of a particular helicity sign is far from overwhelming;
in other words, the minority helicity sense (i.e., sign) has always a 
significant
contribution to the net helicity budget. The helicity imbalance is similar to 
that of quiet Sun areas which are already known to show low degrees of
imbalance \citep[][]{Tziotziou_2014,Tziotziou_2015}. In a separate 
study, it would be interesting to consider the degree 
of compliance of such spatially incoherent distributions of the signed 
components of helicity with the popular scenario that a deep-seated dynamo 
mechanism \citep[e.g. see][and references therein]{Stein_2012,Fan_2021}
sometimes plays a significant role in the generation of AR magnetic flux.

The results from the CB method show moderate qualitative agreement with those
from the flux-integration method, and this is to be expected given the different
nature of the results provided by the two methods: instantaneous values versus
accumulated changes over certain intervals, respectively (see Sect. 4.5 for 
more details). We note in passing that $E_f$ and $H$ around the time of 
AR11890's major flare 5 have also been computed by \citet{Gupta_2021} using a 
finite-volume method.
The overall conclusions for their 10 flare events are different
from ours (they found that $E_f$ and $H$ are not good proxies for the eruptive
potential of the ARs) possibly due to different data and/or method they used.
However, the evolutionary trends they recovered for AR11890's major flare 5 are similar to 
those presented in our Fig. 8(e,f) with the important exceptions that
we deduce peak $E_f$ and $H$ values
that are about 30\% and factor of 3.5, respectively, higher than theirs. The CB method
is supposed to provide lower limits to the $E_f$ and $H$ budgets (see Sect. 
3.1). However, if the NLFF field extrapolations used in the finite-volume
methods converge close to a potential field, then the resulting free energy and
helicity could conceivably be smaller than the ones provided by the CB method.

In both ARs all major eruptive flares occur at times of well-defined 
simultaneous local peaks of both the free magnetic energy and net helicity. The 
discrete, beyond error bars, signature of these changes may reflect 
significant
re-organizations of the AR's magnetic field which is supported by the distinct
appearance of local peaks in the time profiles of the connected flux as well. This
result can be further appreciated if we recall that (1) the instantaneous $E_f$
and $H$ time profiles reflect the net result of the competition between new
free energy and helicity injected into the AR and their removal through flares
($E_f$) and CMEs (both $E_f$ and $H$). (2) There are several studies of major
flares in which the identification of discrete decreases of either the free 
magnetic energy \citep[e.g.][]{Metcalf_2005} or the helicity 
\citep[e.g.][]{Patsourakos_thalis_2016} as a result of the flare was not 
possible beyond uncertainties. 

The occurrence of simultaneous local free magnetic energy and helicity peaks 
during the impulsive phase of the flares is consistent with a paradigm which 
dictates that (1) the prior accumulation of sufficient amounts of free magnetic
energy and helicity is a necessary condition for an AR to erupt, and (2) in the
course of the eruption free energy is released and helicity is bodily removed 
from the AR resulting in the decrease of the budgets of both quantities just 
after the eruption. Furthermore, it is in line with those results that advocate
for a synchronization between the CME acceleration and the impulsive phase of the
flare.

The occurrence of all major flares at times when the ARs contained significant
budgets of both signed components of helicity may appear in favor of the
helicity annihilation mechanism for the onset of solar flares 
\citep[][]{kusano2003,Kusano_2004}. However, we note that in most cases 
the major flares in our ARs occurred around times that the minority 
helicity sense showed small temporal changes.

The results from the computation of the free magnetic energy and net 
helicity changes associated with the eruptive events appear in Tables
2 and 3; the free magnetic energy and net helicity losses ranged from $(0.3-2) 
\times 10^{32}$ erg and $(1.3-20 \times 10^{42}$ Mx$^2$, respectively (their
average values being  $1.02 \times 10^{32}$ erg and $8.68 \times 10^{42}$ Mx$^2$,
respectively). These values are broadly consistent with results from previous 
publications. The percentage losses, associated with the eruptive flares,
in the normalized free magnetic energy were significant, in the range 
$\sim$10-60\%. For the magnetic helicity, changes ranged from $\sim$25\% 
to the removal of the entire excess helicity of the prevailing sign, leading a 
roughly zero net helicity. Such extremely high helicity percentage losses
do not really mean that the active region has turned to potentiality (this 
would be implied if the entire free energy was wiped out, as well, but this is 
not the case). It simply implies that the AR gives out the entire excess 
helicity of one sign and turns to a situation of almost zero net helicity, 
with very significant, but roughly equal and opposite, budgets of both signs. 

Another new result is that we were able to identify the occurrence of the 
eruptive flares at those times when both the free magnetic energy and the net
helicity as well as the prevailining signed component of
helicity local peaks exceed the 2$\sigma$ level of their detrended timeseries. 
Furthermore, there is no eruption when none of these quantities or only
the free magnetic energy or only the helicities
exceeds its 2$\sigma$ level. These results place free energy and helicity on an 
equal footing as far as their role in the initiation of eruptive events. Given
that both ARs possess adequate free energy and helicity budgets 
(that is, higher than those released in a typical eruption) throughout much of
their evolution, this result may plausibly explain the eruption timing.
Clearly, studies of
more ARs are required in order to test whether this threshold represents a 
universal property of eruptive ARs or is peculiar to the ARs studied in 
this paper.

The above result can be reproduced in AR11890 if we use the 
normalized parameters, $E_f/E_{tot}$, and $H/ \Phi_{conn}^2$ or $H_+/
\Phi_{conn}^2$ (see Sect. 4.4) instead of $E_f$ and $H$ or $H_{+}$. 
However, this is not the case for AR11618 because its $E_f/E_{tot}$ curve 
shows two pronounced peaks on November 19 which are not associated with 
eruptions and also its local peak at the time of the second major flare is 
small (see Fig. 7). Therefore the overall potential of this particular set of 
normalized parameters as indicators of the AR approach to eruption 
territory is rather weaker than that of $E_f$ and $H$. Having said that,
the large percentage losses, associated with the eruptive flares, in the 
normalized free magnetic energy and helicity parameters (see Figs 6 and 7 as 
well as Tables 2 and 3) is a particularly notable feature, rarely seen in such 
clarity.

Both the CB and the finite-volume methods cannot provide any density for 
helicity and therefore a detailed picture of the spatial distribution of 
instantaneous helicity is largely unknown. A way to bypass this obstacle is
by employing the concept of relative field line helicity 
\citep[e.g.][]{Yeates_2018,Moraitis_2019}. Although this is a gauge-dependent 
quantity, its first application to solar data \citep[][]{Moraitis_2021} 
shows that its morphology is not sensitive on the gauge used in its
computation, and therefore becomes a promising proxy for the recovery of the 
density of helicity 
in conjunction with the locations of major flare activity. A computation of
the evolution of the field line helicity in the ARs studied here should be
an obvious extension of this work.

\begin{acknowledgements}
We thank the referee for his/her constructive comments.
We thank C.E. Alissandrakis, S. Patsourakos, and K. Moraitis for useful 
discussions. EL acknowledges partial financial support from University of 
Ioannina's internal grants 82985/114841/$\beta$6.$\epsilon$ and 
82985/114842/$\beta$6.$\epsilon$.
\end{acknowledgements}

\bibliographystyle{aa-note} 
\bibliography{ms_accepted}
\end{document}